\documentclass[a4paper,11pt]{article}
\usepackage[utf8]{inputenc}
\usepackage[T1]{fontenc}
\usepackage[english]{babel}
\usepackage[hmargin={32mm,32mm},vmargin={32mm,35mm}]{geometry}
\usepackage{amsmath}
\usepackage{amsfonts}
\usepackage{amssymb}
\usepackage{enumerate}
\usepackage{graphicx}
\usepackage{tocloft}
\usepackage{bbm}
\usepackage{feynmf}
\usepackage{epsfig}
\usepackage{color}
\usepackage{slashed}
\usepackage{xspace}
\usepackage{url}
\usepackage{accents}
\usepackage[makeroom]{cancel}
\usepackage[hidelinks]{hyperref}
\usepackage{afterpage}
\usepackage{cite}
\usepackage{listings}
\usepackage{caption}
\usepackage{epstopdf}
\usepackage{bookmark}

\newcommand{\bra}[1]{\langle #1 |}
\newcommand{\ket}[1]{|#1\rangle}
\newcommand{\braket}[2]{\langle #1 | #2 \rangle}

\newcommand{\bbra}[1]{\bigl\langle #1 \bigr|}
\newcommand{\bket}[1]{\bigl|#1\bigr\rangle}

\newcommand{\be}{\begin{equation}}
\newcommand{\ee}{\end{equation}}
\newcommand{\bea}{\begin{eqnarray}}
\newcommand{\eea}{\end{eqnarray}}
\newcommand{\bal}{\begin{align}}
\newcommand{\eal}{\end{align}}

\newcommand{\eg}{e.g.\@\xspace}
\newcommand{\ie}{i.e.\@\xspace}
\newcommand{\Eq}[1]{Eq.\@\xspace\eqref{#1}}
\newcommand{\Eqs}[1]{Eqs.\@\xspace\eqref{#1}}
\newcommand{\Fig}[1]{Fig.\@\xspace\ref{#1}}

\newcommand{\updown}[2]{^{#1}_{\phantom{#1}#2}}
\newcommand{\downup}[2]{_{#1}^{\phantom{#1}#2}}

\newcommand{\C}{\mathbbm{C}}

\numberwithin{equation}{section}

\newsavebox{\mybox}

\newcommand{\D}[4]{D^{(#1)}_{#2#3}(#4)}
\newcommand{\Db}[4]{D^{(#1)}_{#2#3}\bigl(#4\bigr)}
\newcommand{\CG}[6]{C^{(#1\,#2\,#3)}_{\;#4\,#5\,#6}}
\newcommand{\CGi}[6]{C^{(#1\,#2\,#3)}_{\;#4\,#5\,#6}}
\newcommand{\CGa}[6]{C^{(#1\;#2\;#3)}_{\;#4\;#5\;#6}}

\newcommand{\Tau}[4]{({\tau^{(#1)}_{#2}})_{#3#4}}

\begin{document}

\begin{center}

\Large
\textbf{Operators of quantum-reduced loop gravity \\ from the perspective of full loop quantum gravity}

\vspace{16pt}

\large
Ilkka Mäkinen

\vspace{8pt}

\normalsize
Faculty of Physics, University of Warsaw \\ 
Pasteura 5, 02-093 Warsaw, Poland \\
ilkka.makinen@fuw.edu.pl 

\end{center}

\renewcommand{\abstractname}{\vspace{-\baselineskip}}

\begin{abstract}

\noindent Quantum-reduced loop gravity is a model of loop quantum gravity, which -- from the technical point of view -- is characterized by the remarkably simple form of its basic operators. In this article we examine the operators of the quantum-reduced model from the perspective of full loop quantum gravity. We show that, in spite of their simplicity, the operators of the quantum-reduced model are simply the operators of the full theory acting on states in the Hilbert space of the quantum-reduced model. The passage from the full theory operators to the ''reduced'' operators simply consists of noting that the states of the quantum-reduced model are assumed to carry large spin quantum numbers, and discarding terms which are of lower than leading order in $j$. Our findings clarify the relation between the quantum-reduced model and full loop quantum gravity, and strengthen the technical foundations on which the kinematical structure of the quantum-reduced model is based.

\end{abstract}

\vspace{-8pt}

\enlargethispage{2\baselineskip}

{\let\thefootnote\relax\footnotetext{\hspace{-20.5pt} The work presented in this article is a part of the Polish National Science Centre project Sheng 1, 2018/30/Q/ST2/00811 ''Dynamics and extensions of LQG'', written by Jerzy Lewandowski and Yongge Ma.}}

\section{Introduction}

Quantum-reduced loop gravity is a model proposed by Alesci and Cianfrani \cite{1210, 1301, 1602} in order to address the formidable problem of probing the physical implications of loop quantum gravity \cite{LQG1, LQG2, LQG3, LQG4} -- a problem which has remained a major challenge of loop quantum gravity throughout the three decades that have now passed since the birth of the theory. The quantum-reduced model is based on implementing a gauge fixing to a diagonal spatial metric encoded in a diagonal densitized triad field in the setting of canonical loop quantum gravity. Therefore, even though the early work on quantum-reduced loop gravity was mostly focused on the model's cosmological applications (see \eg \cite{1402, 1410, 1506a}), the model is considerably more general; in principle, it can provide a quantum description of any spacetime represented classically by a diagonal spatial metric. Indeed, the formalism of the quantum-reduced model has recently been extended to spherically symmetric spacetimes \cite{1802, 1807} with the intention of applying the model to study the quantum dynamics of black holes \cite{1904}.

A characteristic feature of quantum-reduced loop gravity is the remarkable simplicity of its operators in comparison with the corresponding operators of full loop quantum gravity. For instance, the reduced volume operator acts diagonally on the natural basis states in the Hilbert space of the quantum-reduced model. This can be contrasted with the situation in the full theory, where even an explicit expression for the matrix elements of the volume operator in the spin network basis is not available, except in certain simple special cases (see \eg \cite{vol1, vol2}). From a practical point of view, this simplicity is a considerable advantage of the quantum-reduced model, as it enables one to explore the model's physical content through concrete calculations, which would be quite intractable within the framework of proper loop quantum gravity.

Accordingly, the central topic of most of the research on quantum-reduced loop gravity has been the physical and phenomenological applications of the model, particularly in the cosmological context \cite{1402, 1410, 1506a, 1604, 1612a, 1709, 1808, 1811, 1901, 1904}. In addition, some work has been devoted to extending the formalism of the quantum-reduced model to include couplings to various matter fields \cite{1506b, 1612b}. While a couple of articles have sought to clarify the relation between quantum-reduced loop gravity and loop quantum cosmology \cite{1410, 1604}, little attention has been paid to the question of investigating the relation between the quantum-reduced model and the full theory of loop quantum gravity. With the exception of the early article \cite{1309}, this question has remained largely unaddressed in the literature of the quantum-reduced model so far.

\enlargethispage{\baselineskip}

The purpose of this article is to illuminate the relation between the operators of the quantum-reduced model and those of full loop quantum gravity. In the standard construction of the quantum-reduced model, the operators of the model are introduced as projections of the corresponding operators of the full theory down to the reduced Hilbert space. However, we will show that these ''reduced'' operators are simply the full-fledged operators of proper loop quantum gravity acting on states in the reduced Hilbert space. More precisely, keeping in mind that the ''reduced spin network states'' of the quantum-reduced model are assumed to carry large spins on each of their edges, the result we will demonstrate is the following: When an operator of full loop quantum gravity, such as the holonomy operator or the volume operator, acts on a state in the reduced Hilbert space, the term of leading order in $j$ reproduces the simple action of the corresponding ''reduced'' operator\footnote{In the case of the holonomy operator, the leading term in $j$ gives a modified form of the reduced holonomy operator, as we will see in section \ref{sec:holonomy}.}. In other words, the discrepancy between the full theory operator and the reduced operator is of lower order in $j$, and is therefore negligible in comparison with the leading term.

This result puts the kinematical framework of the quantum-reduced model on a more solid technical foundation, since it shows that the only genuine technical assumption of the model is the structure of the reduced Hilbert space -- including, in particular, the requirement that the spin quantum number associated to each edge of a ''reduced spin network state'' is large. Once the reduced Hilbert space is given, the (extraordinarily simple) operators of the quantum-reduced model are obtained without introducing any additional assumptions, simply by letting the operators of the full theory act on states in the reduced Hilbert space, and dropping terms which are negligible in the limit of large $j$.

The material in this article is organized as follows. After the present introductory section, we give a brief outline of the kinematical structure of loop quantum gravity in section 2. In section 3 we provide an equally brief overview of the kinematics of the quantum-reduced model, describing the states which form the reduced Hilbert space, and the elementary ''reduced'' operators of the model. In section 4 we present our analysis of the operators of the quantum-reduced model, regarded as operators of the full theory acting on states in the reduced Hilbert space. We will consider the holonomy operator, the flux operator, the volume operator, and a particular version of the Hamiltonian constraint operator, which has been used previously in the literature of the quantum-reduced model. Our conclusions are then given in section 5. The article also contains two appendices, in which we review some useful results from $SU(2)$ representation theory and the quantum theory of angular momentum, and display the solution of a certain technical problem related to extracting the action of the volume operator in the reduced Hilbert space.

\section{Loop quantum gravity}

In this section we give a concise review of the basic kinematical framework of loop quantum gravity. We will describe the kinematical Hilbert space of the theory, and the elementary operators thereon. A complete presentation of the kinematics of loop quantum gravity (see \eg \cite{LQG1, LQG2, LQG3, LQG4}) would go on to introduce the spaces of gauge invariant and diffeomorphism invariant states. However, these spaces do not play any role in the work presented in this article, and we will therefore not discuss them in any detail.

\subsection{The kinematical Hilbert space}\label{sec:Hkin}

The kinematical Hilbert space of loop quantum gravity is the space of so-called cylindrical functions\footnote{More precisely, the kinematical Hilbert space is the completion of the space of cylindrical functions with respect to the scalar product defined by \Eqs{cyl-sp} and \eqref{cyl-sp-diff}.}. A cylindrical function is essentially a (complex-valued) function of the form
\be\label{Psi_G}
\Psi_\Gamma(h_{e_1},\dots,h_{e_N}).
\ee
It is labeled by a graph $\Gamma$, which consists of the edges $e_1,\dots,e_N$. The arguments of the function are $SU(2)$ group elements, one for each edge of the graph. If there is a need to specifically emphasize the graph on which a cylindrical function is defined, the function \eqref{Psi_G} can be said to be cylindrical with respect to the graph $\Gamma$.

The group elements $h_e$ originate from holonomies of the Ashtekar connection in the classical theory, and for this reason they are referred to as holonomies also in the quantum theory. The holonomies satisfy certain algebraic properties, reflecting the classical interpretation of the holonomy as a parallel transport operator. Letting $e^{-1}$ denote the edge $e$ taken with the opposite orientation, we have
\be\label{h-1}
h_{e^{-1}} = h_e^{-1}.
\ee
Furthermore, if $e_1$ and $e_2$ are two edges such that the endpoint of $e_1$ coincides with the beginning point of $e_2$, we have
\be\label{h2h1}
h_{e_2}h_{e_1} = h_{e_2\circ e_1},
\ee
where $e_2\circ e_1$ stands for the edge composed of $e_1$ followed by $e_2$. 

Due to the properties \eqref{h-1} and \eqref{h2h1}, there is a considerable freedom in choosing the graph with respect to which a given cylindrical function is considered to be cylindrical. In particular, any cylindrical function defined on a graph $\Gamma$ can also be viewed as a cylindrical function on any larger graph $\Gamma'$, which contains the graph $\Gamma$ as a subgraph. Letting $e_{N+1},\dots,e_{N'}$ denote the edges of $\Gamma'$ that are not contained in $\Gamma$, the function \eqref{Psi_G} can be trivially rewritten as
\be
\Psi'_{\Gamma'}(h_{e_1},\dots,h_{e_N},h_{e_{N+1}},\dots,h_{e_{N'}}),
\ee
where the function $\Psi'_{\Gamma'}$ is constant with respect to the arguments $h_{e_{N+1}},\dots,h_{e_{N'}}$, and is equal to $\Psi_\Gamma(h_{e_1},\dots,h_{e_N})$ independently of their values.

The observation of the previous paragraph contains the key to defining a scalar product on the space of cylindrical functions. For two functions cylindrical with respect to the same graph $\Gamma$, we may define
\be\label{cyl-sp}
\braket{\Psi_\Gamma}{\Phi_\Gamma} = \int dg_1\,\dots\,dg_N\,\overline{\displaystyle \Psi_\Gamma(g_1,\dots,g_N)} \Phi_\Gamma(g_1,\dots,g_N),
\ee
where $dg$ denotes the Haar measure of $SU(2)$. In order to extend the definition to two functions $\Psi_{\Gamma_1}$ and $\Phi_{\Gamma_2}$, cylindrical with respect to two different graphs $\Gamma_1$ and $\Gamma_2$, we may take any graph $\Gamma_{12}$ that contains both $\Gamma_1$ and $\Gamma_2$ as subgraphs, and view $\Psi_{\Gamma_1}$ and $\Phi_{\Gamma_2}$ as cylindrical functions on $\Gamma_{12}$. The scalar product between the two functions can then be defined as
\be\label{cyl-sp-diff}
\braket{\Psi_{\Gamma_1}}{\Phi_{\Gamma_2}} \equiv \braket{\Psi_{\Gamma_{12}}}{\Phi_{\Gamma_{12}}},
\ee
where the right-hand side is given by \Eq{cyl-sp}. The normalization of the Haar measure guarantees that the value of $\braket{\Psi_{\Gamma_1}}{\Phi_{\Gamma_2}}$ does not depend on how the graph $\Gamma_{12}$ is chosen. The scalar product defined by \Eqs{cyl-sp} and \eqref{cyl-sp-diff} is usually referred to as the Ashtekar--Lewandowski scalar product.

According to the Peter--Weyl theorem, a basis on the space of cylindrical functions can be constructed using the $SU(2)$ representation matrices $\D{j}{m}{n}{h}$. The functions
\be\label{basis}
(\Psi_{\Gamma})^{(j_1\dots j_N)}_{m_1\dots m_N;n_1\dots n_N}(h_{e_1},\dots,h_{e_N}) = \prod_{e\in\Gamma} \D{j_e}{m_e}{n_e}{h_e}
\ee
span the space of functions cylindrical with respect to the graph $\Gamma$, as the quantum numbers $\{j_e\}$, $\{m_e\}$ and $\{n_e\}$ range over all their possible values. The functions \eqref{basis} are orthogonal but not normalized under the scalar product \eqref{cyl-sp}. In order to normalize them, one has to multiply each representation matrix by the factor $\sqrt{d_{j_e}}$.

\subsection{Elementary operators}

The elementary operators of loop quantum gravity are the holonomy and flux operators. The holonomy operator is associated to an edge $e$, and it acts on cylindrical functions by multiplication:
\be\label{Dh*Psi}
\D{j}{m}{n}{h_e}\Psi_\Gamma(h_{e_1},\dots,h_{e_N}).
\ee
The character of the result depends on whether the edge $e$ is contained among the edges of the graph $\Gamma$. If $e$ is not an edge of $\Gamma$, the function \eqref{Dh*Psi} defines a state based on the graph $\Gamma\cup e$; in effect, the action of the holonomy operator has added a new edge to the graph of the state on which it acted. On the other hand, if $e$ coincides with one of the edges of $\Gamma$, the state \eqref{Dh*Psi} is still based on the graph $\Gamma$. In this case, the basic tool for computing the action of the holonomy operator is the Clebsch--Gordan series of $SU(2)$,
\be\label{CGs}
\D{j_1}{m_1}{n_1}{h_e}\D{j_2}{m_2}{n_2}{h_e} = \sum_{j} \CG{j_1}{j_2}{j}{m_1}{m_2}{m_1+m_2}\CG{j_1}{j_2}{j}{n_1}{n_2}{n_1+n_2}\D{j}{m_1+m_2\;}{n_1+n_2}{h_e},
\ee
where $\CG{j_1}{j_2}{j}{m_1}{m_2}{m}$ are the $SU(2)$ Clebsch--Gordan coefficients. (See Appendix \ref{app:SU2} for our notation and conventions regarding the Clebsch--Gordan coefficients and other objects of $SU(2)$ representation theory.) If the orientation of the holonomy operator is opposite to the orientation of the edge on which the operator is acting, one can compute its action by first using the relation
\be
\D{j}{m}{n}{h_e^{-1}} = (-1)^{m-n}\D{j}{-n\;}{-m}{h_e}
\ee
for the matrix elements of the inverse Wigner matrix, and then using \Eq{CGs}.

In order to discuss the flux operator, it is convenient to start by defining a set of auxiliary operators $J_i^{(v,e)}$. Each of these operators carries an $SU(2)$ vector index $i$, and is
labeled by a point $v$ and an edge $e$ such that $v$ is either the beginning or the ending point of $e$. The action of the operator $J_i^{(v,e)}$ on a cylindrical function based on a graph $\Gamma$ is defined to be\footnote{The two cases in \Eq{J*Psi} define the left- and right-invariant vector fields of $SU(2)$.}
\begin{align}
J_i^{(v,e)}&\Psi_\Gamma(h_{e_1},\dots,h_{e_N}) \notag \\
&= \begin{cases} i\dfrac{d}{d\epsilon}\bigg|_{\epsilon=0}\Psi_\Gamma(h_{e_1},\dots,h_{e_k}e^{\epsilon\tau_i},\dots,h_{e_N}) & \text{if $e=e_k$ and $e$ begins at $v$} \\ \vspace{-8pt}  \\
-i\dfrac{d}{d\epsilon}\bigg|_{\epsilon=0}\Psi_\Gamma(h_{e_1},\dots,e^{\epsilon\tau_i}h_{e_k},\dots,h_{e_N}) & \text{if $e=e_k$ and $e$ ends at $v$} \end{cases} \label{J*Psi}
\end{align}
where $\tau_i = -i\sigma_i/2$ are the anti-Hermitian generators of $SU(2)$. If $v$ is not a node of $\Gamma$, or $e$ is not an edge of $\Gamma$, we set $J_i^{(v,e)}\Psi_\Gamma(h_{e_1},\dots,h_{e_N}) = 0$. It is immediate to see that the action of $J_i^{(v,e)}$ on a holonomy is given by
\be\label{J*h s}
J_i^{(v,e)}D^{(j)}(h_e) = iD^{(j)}(h_e)\tau_i^{(j)} \qquad \text{($e$ begins at $v$)}
\ee
and
\be\label{J*h t}
J_i^{(v,e)}D^{(j)}(h_e) = -i\tau_i^{(j)}D^{(j)}(h_e) \qquad \text{($e$ ends at $v$)}
\ee
where $\tau_i^{(j)}$ are the generators of $SU(2)$ in the spin-$j$ representation. (An explicit definition of $\tau_i^{(j)}$ can be read off from \Eqs{tau^j} and \eqref{Jzjm}--\eqref{Jpm} in the Appendix.)

The flux operator $E_i(S)$ associated to a surface $S$ is a quantization of the classical function $\int_S d^2\sigma\,n_aE^a_i$, where $E^a_i$ is the densitized triad field. The flux operator can be expressed in terms of the operator $J_i^{(v,e)}$ as
\be\label{flux}
E_i(S)\Psi_\Gamma(h_{e_1},\dots,h_{e_N}) = 8\pi\beta G\sum_{x\in S}\sum_{\text{$e$ at $x$}} \frac{1}{2}\kappa(S,e) J_i^{(x,e)}\Psi_\Gamma(h_{e_1},\dots,h_{e_N}),
\ee
where $\beta$ is the Barbero--Immirzi parameter, and the geometric factor $\kappa(S,e)$ is
\be
\kappa(S,e) = \begin{cases} +1 & \text{if $e$ lies above $S$} \\
-1 & \text{if $e$ lies below $S$} \\
0 & \text{if $e$ intersects $S$ tangentially or not at all} \end{cases}
\ee
Here ''above'' and ''below'' are understood with respect to the direction defined by the normal vector of the surface. The expression on the right-hand side of \Eq{flux} is well-defined despite the uncountable sum over all the points of $S$, since the sum receives non-vanishing contributions only from the finite number of points at which the edges of the graph $\Gamma$ intersect the surface $S$.

The action of the flux operator on a holonomy can be deduced from \Eqs{J*h s}--\eqref{flux}. For instance, in the case that the edge $e$ lies entirely above the surface $S$ in the sense explained above, we find
\be
E_i(S)D^{(j)}(h_e) = 8\pi\beta G\dfrac{i}{2}D^{(j)}(h_e)\tau_i^{(j)}
\ee
if the beginning point of $e$ lies on $S$, and
\be
E_i(S)D^{(j)}(h_e) = -8\pi\beta G\dfrac{i}{2}\tau_i^{(j)}D^{(j)}(h_e)
\ee
if the endpoint of $e$ lies on $S$. If the surface $S$ intersects the edge $e$ at an interior point, we have
\be
E_i(S)D^{(j)}(h_e) = 8\pi\beta G\nu(S,e)iD^{(j)}(h_{e_2})\tau_i^{(j)}D^{(j)}(h_{e_1}),
\ee
where the factor $\nu(S,e)$ equals $+1$ if the orientation of $e$ agrees with the direction of the normal vector of $S$, and $-1$ if the orientation of the edge is opposite to that of the surface.

\section{The quantum-reduced model}

In this section we introduce the basic kinematical states and elementary operators of quantum-reduced loop gravity, mirroring the outline of the full theory given in the previous section. For the purposes of the present article, it is not necessary to go into the technical details of how the kinematical states of the quantum-reduced model are obtained as the solutions of the corresponding gauge-fixing constraints. We may simply regard the Hilbert space of the quantum-reduced model as a given subspace of the kinematical Hilbert space of the full theory. A discussion of the gauge-fixing procedure which leads to the reduced Hilbert space can be found \eg in the review article \cite{1612a}.

\subsection{The reduced Hilbert space}

The Hilbert space of the quantum-reduced model is constructed by implementing (in the weak sense) certain reduction constraints on the kinematical Hilbert space described in section \ref{sec:Hkin}. These constraints are designed to implement a gauge fixing to a diagonal spatial metric described by a diagonal triad field. The Hilbert space resulting from the reduction is spanned by basis states which have the form \eqref{basis}, and are characterized by the following requirements:
\begin{itemize}
\item The edges of the graph $\Gamma$ are aligned along the $x$-, $y$- and $z$-directions defined by a fiducial background coordinate system.
\item The spin quantum number associated to each edge is large,
\be
j_e\gg 1
\ee
for every edge of the graph.
\item Each edge carries a representation matrix, both of whose magnetic indices take either the maximal or the minimal value (\ie $j_e$ or $-j_e$) with respect to the basis corresponding to the direction of the edge.
\end{itemize}
Let us denote by $\ket{jm}_i$ (where $i=x$, $y$ or $z$) the state which diagonalizes the operators $J^2$ and $J_i$ with eigenvalues $j(j+1)$ and $m$, and introduce the notation
\be
\D{j}{m}{n}{h}_i \equiv {}_i\bra{jm}D^{(j)}(h)\ket{jn}_i
\ee
for the matrix elements of the Wigner matrices in the basis $\ket{jm}_i$. (See section \ref{sec:xystates} of the Appendix for more details on how the states $\ket{jm}_i$ are defined.) Then the wave function of a generic basis state of the reduced Hilbert space has the form
\be\label{basis-r}
\prod_{e\in\Gamma} \D{j_e}{\sigma_ej_e\,}{\sigma_ej_e}{h_e}_{i_e},
\ee
where each $\sigma_e$ is equal to $+1$ or $-1$, and each $i_e$ takes the value $x$, $y$ or $z$, depending on whether the edge $e$ is aligned along the $x$-, $y$- or $z$-direction. 

As a convenient terminology, we will often refer to a state of the form \eqref{basis-r} as a reduced spin network state, and a holonomy of the form $\D{j_e}{\sigma_ej_e\,}{\sigma_ej_e}{h_e}_{i_e}$ as a reduced holonomy. However, the state \eqref{basis-r} is not a spin network state in the sense in which the term is usually understood in loop quantum gravity, namely a basis state of the gauge invariant Hilbert space, in which the representation matrices associated to the edges are contracted with invariant tensors at the nodes of the graph. Indeed, the states \eqref{basis-r} are neither gauge invariant nor diffeomorphism invariant, reflecting the fact that the fundamental assumption of the quantum-reduced model is a restriction to diagonal spatial metrics described by diagonal triads, which breaks both invariance under spatial diffeomorphisms and the internal gauge invariance associated with rotations of the triad.

Let us also emphasize that there are no intertwiners involved in the states \eqref{basis-r}, even though the basis states of the reduced Hilbert space are often (especially in the older literature of the quantum-reduced model) defined by inserting so-called reduced intertwiners at the nodes of the graph. However, as pointed out for the first time in \cite{1612b}, the ''reduced intertwiners'' are simply constant complex numbers multiplying the basis states \eqref{basis-r}. As such, there can be no physically meaningful information contained in them, and they should simply be discarded in order to not needlessly complicate the formalism.

If we keep track of the orientation of the graph on which the state \eqref{basis-r} is defined, then the relation
\be\label{Djj-inv}
\D{j}{j}{j}{h^{-1}} = \D{j}{-j\;}{-j}{h}
\ee
implies that we may restrict ourselves to the case $\sigma_e=+1$ in \Eq{basis-r}, and work with holonomies of the form $\D{j}{j}{j}{h_e}_{i}$ only. Holonomies of the form $\D{j}{-j\;}{-j}{h_e}_i$ do not need to be considered, since \Eq{Djj-inv} shows that a holonomy with magnetic indices $-j,-j$ is equivalent to a holonomy with indices $jj$, and with a reversed orientation of the edge. Alternatively, one could work with both types of holonomies while taking an arbitrary but fixed orientation of the graph. When we come to the analysis presented in section \ref{sec:main}, it is more convenient to take the former point of view, since we can then consider the action of operators only on holonomies of the type $\D{j}{j}{j}{h_e}_{i}$, and do not need to separately discuss the case $\D{j}{-j}{\;-j}{h_e}_{i}$.

\subsection{Reduced operators}\label{sec:red-op}

The elementary operators of the quantum-reduced model are introduced as projections of the corresponding operators of the full theory down to the reduced Hilbert space. As a result of the projection, the magnetic indices of the reduced holonomy operator ${}^R\! D^{(s)}(h_e)$ will be set equal to their maximal or minimal value. The action of the operator is given by the following reduced recoupling rule, which is essentially the multiplication law of the group $U(1)$:
\be\label{Dj+s}
{}^R\!\D{s}{s}{s}{h_e}\D{j}{j}{j}{h_e} = \D{j+s}{j+s\;}{j+s}{h_e}
\ee
and
\be\label{Dj-s}
{}^R\!\D{s}{-s}{\;-s}{h_e}\D{j}{j}{j}{h_e} = \D{j-s}{j-s\;}{j-s}{h_e}
\ee
and similarly for the case where the operator acts on a reduced holonomy carrying magnetic indices $-j$, $-j$. The multiplication law given by \Eqs{Dj+s} and \eqref{Dj-s} was introduced in \cite{1506a} to replace the somewhat different form of the reduced recoupling rule proposed originally in \cite{1301}.

The flux operators of the quantum-reduced model are associated only to surfaces dual to the coordinate directions of the fiducial coordinate system, \ie to surfaces $S_k$ such that the fiducial coordinate $x^k$ is constant on $S_k$. The reduced flux operator ${}^R\! E_i(S_k)$ is non-vanishing only if $i=k$. Moreover, when acting on a reduced holonomy associated to an edge $e$, the result vanishes unless the surface of the flux operator is dual to the direction of the edge. In the non-vanishing case, the reduced flux operator acts diagonally, picking out the magnetic index of the reduced holonomy on which it is acting. The action of the reduced flux operator is therefore summarized by the equations
\be\label{flux-red-diag}
{}^R\! E_i(S_i) \D{j}{\sigma j\,}{\sigma j}{h_e}_i = (8\pi\beta G)\sigma j\D{j}{\sigma j\,}{\sigma j}{h_e}_i
\ee
(assuming there is an intersection between the edge $e$ and the surface $S_i$), and
\be\label{flux-red-off}
{}^R\! E_i(S_k) \D{j}{\sigma j\,}{\sigma j}{h_e}_l = 0 \qquad \text{if $i\neq k$ or $k\neq l$}.
\ee
The diagonal action of the reduced flux operator implies that operators which are constructed out of the flux operator are extremely simple in the quantum-reduced model. In particular, the reduced volume operator acts diagonally on the basis states \eqref{basis-r}. This can be contrasted with the situation in proper loop quantum gravity, where not even an explicit expression for the matrix elements of the volume operator in the spin network basis is available, except in certain special cases. Indeed, a characteristic feature of the quantum-reduced model is the considerable simplicity of its operators in comparison to the corresponding operators of the full theory.

\section{Reduced operators from the perspective of full LQG}\label{sec:main}

We now move on to the main topic of this article, namely a demonstration of how the operators of the quantum-reduced model are related to the corresponding operators in full loop quantum gravity. In addition to the holonomy and flux operators, we will also consider the volume operator, and the Euclidean part of Thiemann's Hamiltonian in a particular regularization, which has been used in previous works to study the dynamics of the quantum-reduced model.

The general picture that emerges from our calculations is the following: As we let the operators of the full theory act on holonomies of the form $\D{j}{j}{j}{h_e}$, recalling that the value of $j$ is assumed to be large, we find that the term of highest order in $j$ resulting from the action of a given operator agrees with the action of the corresponding quantum-reduced operator. The discrepancy between the full theory operator and the reduced operator is of lower order in $j$ compared to the leading term. (In the case of the holonomy operator, the multiplication law \eqref{Dj+s}--\eqref{Dj-s} of the reduced holonomy operator is recovered only when the holonomy operator carries spin $1/2$. For higher spins we obtain a somewhat modified version of the reduced recoupling rule.)

The results found in this section show that the ''reduced'' operators of the quantum-reduced model should not be thought of as objects unrelated to the operators of the full theory, whose action in the reduced Hilbert space is simply postulated -- and not even as the full theory operators projected down to the Hilbert space of the quantum-reduced model. Instead, they are simply the proper operators of the full theory acting on states in the reduced Hilbert space (which is a genuine subspace of the kinematical Hilbert space of the full theory). The only ''reduction'' of the operators that actually takes place merely amounts to keeping in mind that one is working in the limit of large spins, and discarding terms which are of lower than leading order in $j$.

\subsection{Holonomy operator}\label{sec:holonomy}

We begin our discussion with an analysis of the holonomy operator. We must consider the action of the operator $\D{s}{m}{n}{h_e}$ on a holonomy of the ''reduced'' form $\D{j}{j}{j}{h_e}$. In addition to assuming $j\gg 1$, we will also assume that $s\ll j$, since if $s$ were of the same order of magnitude as $j$, the action of the holonomy operator would not necessarily preserve the requirement of the value of $j$ being large.

Before taking on the general problem of a holonomy operator carrying an arbitrary spin $s$, we will discuss separately the two simplest examples, namely $s=1/2$ and $s=1$. The first example illustrates the mechanism through which the recoupling rule of the reduced holonomy operator is reproduced from the action of the holonomy operator of the full theory. The second example shows that when $s>1/2$, we should expect to recover a slightly modified form of the multiplication law for reduced holonomies. Our findings from the two examples will help us to anticipate the result of the calculation in the general case.

\subsubsection*{Example: Spin 1/2}

Let us first study the action of a holonomy operator carrying spin 1/2. This is given by \Eq{CGs} as
\be
\D{1/2}{A}{B}{h_e}\D{j}{j}{j}{h_e} = \sum_k \CGa{j}{1/2}{k}{j}{A}{j+A}\CGa{j}{1/2}{k}{j}{B}{j+B}\D{k}{j+A\;}{j+B}{h_e},
\ee
where the sum over $k$ runs over the two values $k=j\pm 1/2$.

We may consider the different possible values of the indices $A$ and $B$ case by case\footnote{In what follows, we will use ''$+$'' and ''$-$'' as a shorthand for the two possible values $+1/2$ and $-1/2$ of the indices $A$ and $B$.}. When $A=B=+$, we immediately obtain
\be \label{h++}
\D{1/2}{+}{+}{h_e}\D{j}{j}{j}{h_e} = \D{j+1/2}{\;j+1/2\;}{j+1/2}{h_e},
\ee
since $k=j+1/2$ is the only value of the total spin that is consistent with the value $j+1/2$ of the total magnetic number. In this case the action of the holonomy operator in the quantum-reduced model is reproduced exactly.

In the case $A=B=-$, we have
\begin{align}
\D{1/2}{-}{-}{h_e}\D{j}{j}{j}{h_e} &= \Bigl(\CGa{j}{1/2}{j+1/2}{j}{1/2}{j-1/2}\Bigr)^2\D{j+1/2}{\; j-1/2\;}{j-1/2}{h_e} \notag \\
&+ \Bigl(\CGa{j}{1/2}{j-1/2}{j}{1/2}{j-1/2}\Bigr)^2\D{j-1/2}{\; j-1/2\;}{j-1/2}{h_e}. \label{h--}
\end{align}
At a first sight this does not seem to be compatible with the recoupling rule of the quantum-reduced model, since the first term on the right-hand side contains a holonomy whose magnetic indices are not equal to their maximal (nor minimal) value. However, noting that the Clebsch--Gordan coefficients involved in the above equation are given by
\be
\CGa{j}{1/2}{j+1/2}{j}{1/2}{j-1/2} = \frac{1}{\sqrt{2j+1}}, \qquad \CGa{j}{1/2}{j-1/2}{j}{1/2}{j-1/2} = \sqrt{\frac{2j}{2j+1}},
\ee
and recalling that the spin $j$ is assumed to be large, we see that \Eq{h--} reduces to
\be\label{h-- final}
\D{1/2}{-}{-}{h_e}\D{j}{j}{j}{h_e} = \D{j-1/2}{\;j-1/2\;}{j-1/2}{h_e} + {\cal O}\biggl(\frac{1}{j}\biggr).
\ee
In other words, we find that even though the action of the reduced holonomy operator is not recovered exactly, the discrepancy is of subleading order in $j$, and therefore becomes negligible in the limit of large $j$.

When the indices $A$ and $B$ are not equal to each other, the action of the operator $\D{1/2}{A}{B}{h_e}$ on the state $\D{j}{j}{j}{h_e}$ cannot produce a holonomy in which both magnetic indices are equal to the maximal (or minimal) value. However, in this case the action of the operator gives a result which is entirely of lower order in $j$, in comparison to the leading terms in \Eqs{h++} and \eqref{h-- final}:
\be
\D{1/2}{+}{-}{h_e}\D{j}{j}{j}{h_e} = \CGa{j}{1/2}{j+1/2}{j}{1/2}{j+1/2}\CGa{j}{1/2}{j+1/2}{j}{1/2}{j-1/2}\D{j+1/2}{\;j+1/2\;}{j-1/2}{h_e} = {\cal O}\biggl(\frac{1}{\sqrt j}\biggr)
\ee
and similarly for the operator $\D{1/2}{-}{+}{h_e}$.

Hence the conclusion in the example at hand is that the multiplication law of the reduced holonomy operator is reproduced approximately in the full theory; the approximation amounts to remembering that one is working with large values of the spin $j$, and neglecting terms which are of lower than leading order in $j$.

\subsubsection*{Example: Spin 1}

Before moving on to discuss the general case, let us take a look at the example of a holonomy operator carrying spin 1, since this example will reveal a new feature which was not encountered in the case of a holonomy operator in the fundamental representation.

The action of a spin-1 holonomy operator on the state $\D{j}{j}{j}{h_e}$ is given by
\be
\D{1}{m}{n}{h_e}\D{j}{j}{j}{h_e} = \sum_k \CGa{j}{1}{k}{j}{m}{j+m}\CGa{j}{1}{k}{j}{n}{j+n}\D{k}{j+m\;}{j+n}{h_e},
\ee
where the sum over $k$ now ranges through the values $k=j-1$, $j$ and $j+1$. In the case $m=n=1$ we find, as in the previous example,
\be
\D{1}{1}{1}{h_e}\D{j}{j}{j}{h_e} = \D{j+1}{\;j+1\;}{j+1}{h_e}.
\ee
When $m=n=-1$, we obtain three terms from the sum over $k$:
\begin{align}
\D{1}{-1\;}{-1}{h_e}\D{j}{j}{j}{h_e} &= \Bigl(\CGa{j}{1}{j+1}{j}{-1}{j-1}\Bigr)^2\D{j+1}{\;j-1\;}{j-1}{h_e} \notag \\
&+ \Bigl(\CGa{j}{1}{j}{j}{-1}{j-1}\Bigr)^2\D{j}{\;j-1\;}{j-1}{h_e} + \Bigl(\CGa{j}{1}{j-1}{j}{-1}{j-1}\Bigr)^2\D{j-1}{\;j-1\;}{j-1}{h_e}.\label{D--}
\end{align}
Inserting the values of the relevant Clebsch--Gordan coefficients,
\begin{align}
\CGa{j}{1}{j+1}{j}{-1}{j-1} &= \frac{1}{\sqrt{(j+1)(2j+1)}} \\
\CGa{j}{1}{j}{j}{-1}{j-1} &= \frac{1}{\sqrt{j+1}} \\
\CGa{j}{1}{j-1}{j}{-1}{j-1} &= \sqrt{\frac{2j-1}{2j+1}}
\end{align}
we find that \Eq{D--} becomes
\be
\D{1}{-1\;}{-1}{h_e}\D{j}{j}{j}{h_e} = \D{j-1}{\;j-1\;}{j-1}{h_e} + {\cal O}\biggl(\frac{1}{j}\biggr),
\ee
again in full analogy with the spin-1/2 example.

The new feature is encountered when we consider the action of a holonomy operator with $m=n=0$. In this case we have
\be
\D{1}{0}{0}{h_e}\D{j}{j}{j}{h_e} = \Bigl(\CGa{j}{1}{j+1}{j}{0}{j}\Bigr)^2\D{j+1}{j}{j}{h_e} + \Bigl(\CGa{j}{1}{j}{j}{0}{j}\Bigr)^2\D{j}{j}{j}{h_e},
\ee
where the Clebsch--Gordan coefficients are given by
\be
\CGa{j}{1}{j+1}{j}{0}{j} = \frac{1}{\sqrt{j+1}}, \qquad \CGa{j}{1}{j}{j}{0}{j} = \sqrt{\frac{j}{j+1}}.
\ee
Hence we obtain
\be
\D{1}{0}{0}{h_e}\D{j}{j}{j}{h_e} = \D{j}{j}{j}{h_e} + {\cal O}\biggl(\frac{1}{j}\biggr).
\ee
This shows that the operator $\D{1}{0}{0}{h_e}$ acts in an appropriate way as a ''quantum-reduced'' operator, adding 0 units of spin to the reduced holonomy on which it is acting. This is a departure from the usual formulation of the quantum-reduced model, in which only $\D{1}{1}{1}{h_e}$ and $\D{1}{-1\;}{-1}{h_e}$ would be considered as valid reduced operators, with the operator $\D{1}{0}{0}{h_e}$ not entering the formulation of the model.

Finally, when $m\neq n$, the action of the operator $\D{1}{m}{n}{h_e}$ again produces a result of subleading order in $j$. For example, we find
\be
\D{1}{1}{0}{h_e}\D{j}{j}{j}{h_e} = {\cal O}\biggl(\frac{1}{\sqrt j}\biggr), \qquad \D{1}{0\;}{-1}{h_e}\D{j}{j}{j}{h_e} = {\cal O}\biggl(\frac{1}{\sqrt j}\biggr)
\ee
and
\be
\D{1}{1\;}{-1}{h_e}\D{j}{j}{j}{h_e} = {\cal O}\biggl(\frac{1}{j}\biggr)
\ee
and so on.

\subsubsection*{The general case}

We now proceed to consider the general problem of the operator
\be
\D{s}{m}{n}{h_e}
\ee
acting on the state
\be
\D{j}{j}{j}{h_e}
\ee
assuming that $s\ll j$. Based on our findings in the two examples discussed above, we may anticipate the result of the calculation in the general case. We expect to find\footnote{The index $m$ is not summed over in \Eqs{expectation1} and \eqref{DmmDjj}.}
\be\label{expectation1}
\D{s}{m}{m}{h_e}\D{j}{j}{j}{h_e} = \D{j+m}{\;j+m\;}{j+m}{h_e} + {\cal O}\biggl(\frac{1}{j}\biggr)
\ee
and
\be\label{expectation2}
\D{s}{m}{n}{h_e}\D{j}{j}{j}{h_e} = {\cal O}\biggl(\frac{1}{\sqrt j}\biggr)
\ee
whenever $m\neq n$. 

Since
\be\label{DmmDjj}
\D{s}{m}{m}{h_e}\D{j}{j}{j}{h_e} = \sum_k \Bigl(\CGa{j}{s}{k}{j}{m}{j+m}\Bigr)^2\D{k}{j+m\;}{j+m}{h_e},
\ee
the validity of \Eq{expectation1} clearly hinges on the value of the Clebsch--Gordan coefficient $\CGa{j}{s}{j+m}{j}{m}{j+m}$. Our strategy for proving \Eq{expectation1} is to show that
\be\label{expectation3}
\CGa{j}{s}{j+m}{j}{m}{j+m} = 1 + {\cal O}\biggl(\frac{1}{j}\biggr).
\ee
If \Eq{expectation3} holds, the completeness relation
\be
1 = \sum_k \Bigl(\CGa{j}{s}{k}{j}{m}{j+m}\Bigr)^2 = \Bigl(\CGa{j}{s}{j+m}{j}{m}{j+m}\Bigr)^2 + \sum_{k\neq j+m} \Bigl(\CGa{j}{s}{k}{j}{m}{j+m}\Bigr)^2
\ee
then implies that the coefficients $\CGa{j}{s}{k}{j}{m}{j+m}$ with $k\neq j+m$ are of lower order in $j$:
\be\label{expectation4}
\CGa{j}{s}{k}{j}{m}{j+m} = {\cal O}\biggl(\frac{1}{\sqrt j}\biggr) \qquad (k\neq j+m).
\ee
From this it follows that the terms with $k\neq j+m$ in \Eq{DmmDjj} are of order $1/j$, so \Eqs{expectation3} and \eqref{expectation4} are sufficient to ensure that \Eq{expectation1} holds. \Eq{expectation4} also guarantees the validity of \Eq{expectation2}, since each term on the right-hand side of
\be\label{DmnDjj}
\D{s}{m}{n}{h}\D{j}{j}{j}{h} = \sum_k \CGa{j}{s}{k}{j}{m}{j+m}\CGa{j}{s}{k}{j}{n}{j+n}\D{k}{j+m\;}{j+n}{h}
\ee
contains at least one coefficient of the form \eqref{expectation4}.

It now remains to verify the crucial equation \eqref{expectation3}. The important Clebsch--Gordan coefficient has the relatively simple explicit form
\be\label{importantC}
\CGa{j}{s}{j+m}{j}{m}{j+m} = \sqrt{\frac{(2j)!(2j+2m+1)!}{(2j-s+m)!(2j+s+m+1)!}}.
\ee
For large values of $j$, the factorials can be approximated using Stirling's formula. We consider the logarithm of the number under the square root,
\be\label{log}
\ln(2j)! + \ln(2j+2m+1)! - \ln(2j-s+m)! - \ln(2j+s+m+1)!
\ee
and apply Stirling's approximation in the form
\be\label{stirling}
\ln N! = N\ln N - N + \frac{1}{2}\ln 2\pi N + \frac{1}{12N} + {\cal O}\biggl(\frac{1}{N^3}\biggr),
\ee
where we have included all the terms which can in principle lead to contributions of order $1/j$ or higher. When \Eq{stirling} is used to approximate the logarithms in \eqref{log}, the terms of order $N\ln N$ give
\begin{align}
&2j \ln 2j + (2j+2m+1)\ln(2j+2m+1) \notag \\
&- (2j-s+m)\ln(2j-s+m) - (2j+s+m+1)\ln(2j+s+m+1).\label{NlnN}
\end{align}
We now expand the logarithms as
\be
\ln(2j+x) = \ln 2j + \ln\biggl(1+\frac{x}{2j}\biggr) = \ln 2j + \frac{x}{2j} - \frac{x^2}{4j^2} + {\cal O}\biggl(\frac{1}{j^3}\biggr),
\ee
and find that \eqref{NlnN} reduces to
\be\label{log2}
-\frac{s(s+1)-m(m+1)}{2j} + {\cal O}\biggl(\frac{1}{j^2}\biggr).
\ee
As to the remaining terms that result when \Eq{stirling} is applied to \Eq{log}, the terms linear in $N$ immediately sum up to zero, while a short calculation shows that the terms proportional to $\ln N$ and $1/N$ give contributions of order $1/j^2$. Hence the entire contribution at order $1/j$ is that given by \eqref{log2}.

Recalling that \eqref{log2} is an approximation for the logarithm of the number under the square root in \Eq{importantC}, we have found
\be
\ln \CGa{j}{s}{j+m}{j}{m}{j+m} = -\frac{s(s+1)-m(m+1)}{4j} + {\cal O}\biggl(\frac{1}{j^2}\biggr).
\ee
For the Clebsch--Gordan coefficient itself, this implies
\be
\CGa{j}{s}{j+m}{j}{m}{j+m} = 1 - \frac{s(s+1)-m(m+1)}{4j} + {\cal O}\biggl(\frac{1}{j^2}\biggr),
\ee
showing that the coefficient indeed has the form \eqref{expectation3}, and hence confirming that the result anticipated in \Eqs{expectation1} and \eqref{expectation2} is valid.

To summarize our discussion of the holonomy operator, we have shown that the action of the operator $\D{1/2}{m}{n}{h_e}$ on the state $\D{j}{j}{j}{h_e}$ reproduces the multiplication law of the reduced holonomy operator in the quantum-reduced model, up to terms of subleading order in $j$. In the case of a holonomy operator carrying a spin higher than $1/2$, we discovered a modified form of the reduced recoupling rule. Under the modified recoupling rule given by \Eqs{expectation1} and \eqref{expectation2}, all the diagonal components of the operator $\D{s}{m}{n}{h_e}$ (and not only the components labeled by $m=n=\pm s$) act as valid ''quantum-reduced'' operators. Our findings therefore suggest that the label $s$ in the quantum-reduced multiplication law of \Eqs{Dj+s} and \eqref{Dj-s} should not be interpreted as the spin carried by the operator $\D{s}{s}{s}{h_e}$, but rather as the magnetic quantum number of the operator $\D{l}{s}{s}{h_e}$ (keeping in mind that the leading term in the action of the latter operator on the state $\D{j}{j}{j}{h_e}$ is independent of the spin $l$, and is entirely determined by the magnetic number $s$, as long as the assumption $l\ll j$ is satisfied).

Before proceeding to consider the flux operator, let us briefly comment on the normalization of the states used in the above calculations. For simplicity, we have chosen to work with the unnormalized basis states $D^{(j)}_{jj}(h_e)$. However, the results established in this section are not sensitive to this choice. While equations such as \Eq{expectation1} involve different basis states whose norms are not exactly equal to each other, all the states entering the equation (including those contained in the subleading terms) have the same norm at leading order in $j$. Hence the equality between the leading terms would continue to be valid in the same form, even if we restored the correct normalization of the basis states. (On the other hand, if we were interested in finding the precise form of the lower-order correction terms, then it would be important to work with normalized basis states.) 

\subsection{Flux operator}

In order to analyze the action of the flux operator on the reduced Hilbert space, we begin by considering the operator $J_i^{(v,e)}$ defined by \Eq{J*Psi}. Taking a reduced holonomy $\D{j}{j}{j}{h_e}$  of an edge $e$ aligned in the $z$-direction, and assuming that $v$ is the beginning point of the edge $e$, the action of $J_i^{(v,e)}$ gives
\be\label{J_i Djj}
J_i^{(v,e)}\D{j}{j}{j}{h_e} = i\D{j}{j}{m}{h_e}\Tau{j}{i}{m}{j}.
\ee
The matrix elements of the generators entering the above equation can be read off from \Eqs{Jx*z}--\eqref{Jz*z}. We have
\be
\Tau{j}{x}{m}{j} = -i\sqrt{\frac{j}{2}}\delta_{m,j-1}, \qquad \Tau{j}{y}{m}{j} = \sqrt{\frac{j}{2}}\delta_{m,j-1}, \qquad \Tau{j}{z}{m}{j} = -ij\delta_{mj},
\ee
leading to
\be
J_z^{(v,e)}\D{j}{j}{j}{h_e} = j\D{j}{j}{j}{h_e}
\ee
and
\be
J_x^{(v,e)}\D{j}{j}{j}{h_e} = -i\sqrt{\frac{j}{2}}\D{j}{j\;}{j-1}{h_e}, \qquad J_y^{(v,e)}\D{j}{j}{j}{h_e} = \sqrt{\frac{j}{2}}\D{j}{j\;}{j-1}{h_e}.
\ee
The calculation is easily generalized to the case where $v$ is the endpoint of the edge $e$, or where $e$ is oriented along the $x$- or the $y$-direction. When the index $i$ of the operator $J_i^{(v,e)}$ matches the direction of the edge $e$, we obtain
\be\label{Ji*Di}
J_i^{(v,e)}\D{j}{j}{j}{h_e}_i = \pm j\D{j}{j}{j}{h_e}_i,
\ee
where the sign is $+$ if $e$ begins from $v$, and $-$ if $e$ ends at $v$. When the operator $J_i^{(v,e)}$ acts on an edge aligned in a direction different from $i$, we get
\be\label{Ji*Dk}
J_i^{(v,e)}\D{j}{j}{j}{h_e}_k = {\cal O}\bigl(\sqrt j\bigr) \qquad (i\neq k).
\ee
This result has the same general structure as we found in the case of the holonomy operator. When the operator $J^{(v,e)}$ acts on the reduced holonomy $\D{j}{j}{j}{h_e}_i$, the contribution of highest order in $j$ is given by the component $J_i^{(v,e)}$, whose action preserves the form of the reduced holonomy. The action of the components $J_k^{(v,e)}$ with $k\neq i$ gives a result in which the holonomy is not of the appropriate ''reduced'' form (\ie the magnetic indices of the holonomy are not both equal to the maximal or the minimal value). However, the contribution of the $k\neq i$ components is of lower order in $j$ compared to that of $J_i^{(v,e)}$.

Let us then move on to consider the flux operator $E_i(S_k)$ where, according to the discussion of the reduced flux operator in section \ref{sec:red-op}, the surface $S_k$ lies in the plane $x^k={\rm const.}$ defined by the fiducial background coordinate system. When the operator $E_i(S_k)$ acts on the reduced holonomy $\D{j}{j}{j}{h_e}_l$, the geometric factor $\kappa(S,e)$ in \Eq{flux} implies that the result can be non-vanishing only if $k=l$, since only in this case there can be a transversal intersection between the edge $e$ and the surface $S_k$. Assuming further that the surface intersects the edge at one of its endpoints, the action of the flux operator gives
\be\label{flux-end}
E_i(S_k)\D{j}{j}{j}{h_e}_k = (8\pi\beta G)\frac{1}{2}\kappa(S_k,e)J_i^{(v,e)}\D{j}{j}{j}{h_e}_k.
\ee
With the help of \Eqs{Ji*Di} and \eqref{Ji*Dk}, we see that if the index of the flux operator agrees with the direction of the edge $e$, we get
\be\label{flux-diag}
E_i(S_i)\D{j}{j}{j}{h_e}_i = \pm(8\pi\beta G)\frac{1}{2}j\D{j}{j}{j}{h_e}_i,
\ee
while if the index does not match the orientation of $e$, we have
\be\label{flux-off}
E_i(S_k)\D{j}{j}{j}{h_e}_k = {\cal O}\bigl(\sqrt j\bigr) \qquad (i\neq k).
\ee
Up to the factor $\pm 1/2$, \Eqs{flux-diag} and \eqref{flux-off} agree with the action of the reduced flux operator given by \Eqs{flux-red-diag} and \eqref{flux-red-off}, provided that we neglect the contribution of order $\sqrt j$ in comparison with the term of order $j$.

Note, however, that the action of the reduced flux operator is correctly recovered only when the intersection between the edge and the surface is the beginning or ending point of the edge. If the surface intersects the edge at an interior point, we obtain, instead of \Eq{flux-end},
\be\label{flux-int}
E_i(S_k)\D{j}{j}{j}{h_e}_k = 8\pi\beta G\nu(S^k,e)i\D{j}{j}{m}{h_{e_2}}_k (\tau_i^{(j)})^k_{mn}\D{j}{n}{j}{h_{e_1}}_k,
\ee
where $e_1$ and $e_2$ are the two segments into which the edge $e$ is divided by the surface $S_k$, and $(\tau_i^{(j)})^k_{mn}$ denotes the matrix elements of the generator $\tau_i^{(j)}$ in the basis $\ket{jm}_k$. In general there is no reason why the expression \eqref{flux-int} should reduce to a simpler form, even in the limit of large $j$, since all the matrix elements of the generator $\tau_i^{(j)}$ are involved in it, and not only those in which one index is equal to $j$.

On the other hand, most of the operators one is usually dealing with in loop quantum gravity -- for instance, the volume operator discussed in the following section -- can be formulated directly in terms of the operator $J_i^{(v,e)}$, without having to make any explicit reference to the flux operator. On account of this, it does not seem to be a very serious problem that the quantum-reduced form of the flux operator is not valid in complete generality; it is more important that we have established the relations \eqref{Ji*Di} and \eqref{Ji*Dk} for the operator $J_i^{(v,e)}$.

\subsection{Volume operator}\label{sec:volume}

The volume operator in loop quantum gravity \cite{volume} (restricted to a single node $v$ of a cylindrical function) has the form
\be
V_v = \sqrt{|q_v|},
\ee
where
\be\label{q_v}
q_v = \frac{1}{48}\sum_{\substack{e_I,e_J,e_K \\ \text{at $v$}}} \epsilon(e_I,e_J,e_K) \epsilon^{ijk} J_i^{(v,e_I)}J_j^{(v,e_J)}J_k^{(v,e_K)}
\ee
and the orientation factor $\epsilon(e_I,e_J,e_K)$ is equal to $+1$, $-1$ or $0$, depending on whether the triple of vectors defined by the outgoing tangent directions of the edges $e_I$, $e_J$ and $e_K$ is positively oriented, negatively oriented or not linearly independent. As noted in the previous section, the volume operator is expressed entirely in terms of the operators $J_i^{(v,e)}$, and the flux operator is not directly involved in its definition.

\begin{figure}[t]
	\centering
		\includegraphics[width=0.35\textwidth]{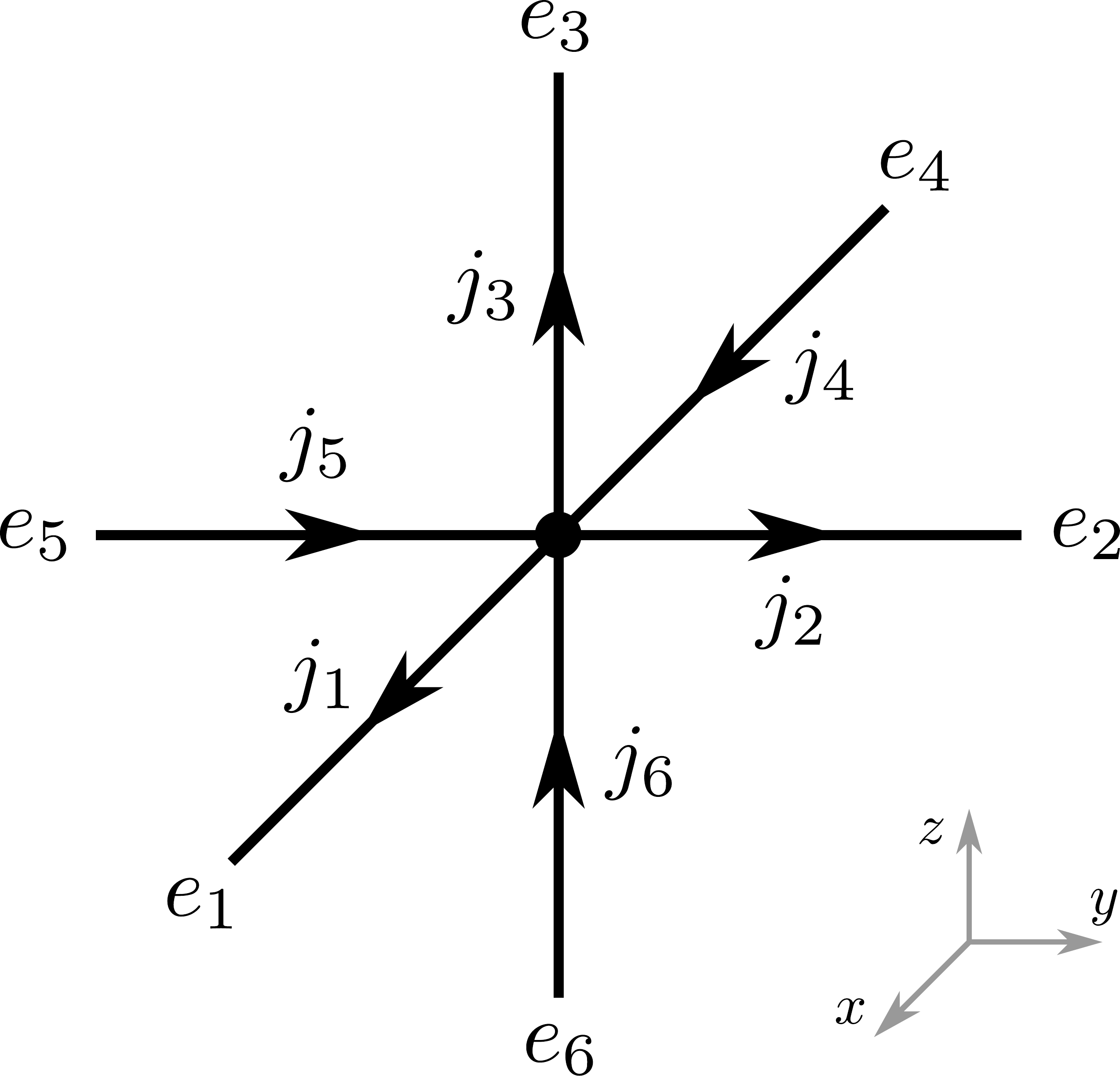}
		\caption{A six-valent node of a reduced spin network state.}
		\label{fig:6node}
\end{figure}

We will study the action of the volume operator on a generic six-valent node of a reduced spin network state. The wave function associated to the node is
\be\label{6-node}
\D{j_1}{j_1}{j_1}{h_{e_1}}_x\,\D{j_2}{j_2}{j_2}{h_{e_2}}_y\,\D{j_3}{j_3}{j_3}{h_{e_3}}_z\,\D{j_4}{j_4}{j_4}{h_{e_4}}_x\,\D{j_5}{j_5}{j_5}{h_{e_5}}_y\,\D{j_6}{j_6}{j_6}{h_{e_6}}_z,
\ee
and we assume that the edges belonging to the node are oriented as shown in \Fig{fig:6node}. Expanding the sum over edges in \Eq{q_v}, we see that when acting on the cuboidal six-valent node, the operator $q_v$ takes the form
\be\label{q_v 6}
q_v = \frac{1}{8}\epsilon^{ijk}\bigl(J_i^{(v,e_1)}-J_i^{(v,e_4)}\bigr)\bigl(J_j^{(v,e_2)}-J_j^{(v,e_5)}\bigr)\bigl(J_k^{(v,e_3)}-J_k^{(v,e_6)}\bigr).
\ee
Recalling \Eqs{Ji*Di} and \eqref{Ji*Dk}, it is immediate to calculate the action of the operator \eqref{q_v 6} on the state \eqref{6-node}, up to terms of lower order in $j$. The leading term is obtained when the indices $i$, $j$ and $k$ in \Eq{q_v} take respectively the values $x$, $y$ and $z$. This term is
\begin{align}
\bigl(J_x^{(v,e_1)}-J_x^{(v,e_4)}\bigr)\bigl(J_y^{(v,e_2)}-{}&{}J_y^{(v,e_5)}\bigr)\bigl(J_z^{(v,e_3)}-J_z^{(v,e_6)}\bigr)\D{j_1}{j_1}{j_1}{h_{e_1}}_x \cdots \D{j_6}{j_6}{j_6}{h_{e_6}}_z \notag \\
&\quad=(j_1+j_4)(j_2+j_5)(j_3+j_6)\D{j_1}{j_1}{j_1}{h_{e_1}}_x \cdots \D{j_6}{j_6}{j_6}{h_{e_6}}_z
\end{align}
(Note that the $+$ sign in \Eq{Ji*Di} applies to the operators associated to the edges $e_1$, $e_2$ and $e_3$, while the $-$ sign applies to the operators acting on the edges $e_4$, $e_5$ and $e_6$.) The remaining terms, in which the triple $(i,j,k)$ is not equal to $(x,y,z)$, are of at least one order of magnitude lower in $j$, since each of these terms contains at least two instances of the operator $J_i^{(v,e)}$ acting on the holonomy of an edge which is not oriented along the $i$-direction. For example, when $(i,j,k) = (x,z,y)$, we get
\be
\underbrace{\bigl(J_x^{(v,e_1)}-J_x^{(v,e_4)}\bigr)}_{{\cal O}(j)}\underbrace{\bigl(J_z^{(v,e_2)}-J_z^{(v,e_5)}\bigr)}_{{\cal O}(\sqrt j)}\underbrace{\bigl(J_y^{(v,e_3)}-J_y^{(v,e_6)}\bigr)}_{{\cal O}(\sqrt j)}\D{j_1}{j_1}{j_1}{h_{e_1}}_x \cdots \D{j_6}{j_6}{j_6}{h_{e_6}}_z = {\cal O}(j^2).
\ee
Hence we find that at leading order in $j$, the action of the operator $q_v$ on the reduced spin network node is diagonal:
\begin{align}
q_v\D{j_1}{j_1}{j_1}{h_{e_1}}_x &\cdots \D{j_6}{j_6}{j_6}{h_{e_6}}_z \notag \\
&= \frac{1}{8}(j_1+j_4)(j_2+j_5)(j_3+j_6) \D{j_1}{j_1}{j_1}{h_{e_1}}_x \cdots \D{j_6}{j_6}{j_6}{h_{e_6}}_z + {\cal O}(j^2). \label{q_v*6node}
\end{align}
However, it does not seem immediately obvious whether we can conclude from this that the operator $V_v = \sqrt{|q_v|}$ enjoys the same property. The answer to this question turns out to be in the affirmative, and is based on the observation that since the off-diagonal terms generated by the action of $q_v$ on the state \eqref{6-node} are of subleading order in $j$ compared to the diagonal term, the action of the square root $\sqrt{|q_v|}$ on \eqref{6-node} can be accessed using standard perturbation theory familiar from quantum mechanics, treating the off-diagonal terms as a perturbation over the diagonal term. The analysis is presented in detail in Appendix \ref{app:sqrt}. The conclusion is that the action of the volume operator on the reduced spin network node is indeed diagonal at leading order in $j$, and is given by
\begin{align}
V_v&\D{j_1}{j_1}{j_1}{h_{e_1}}_x \cdots \D{j_6}{j_6}{j_6}{h_{e_6}}_z \notag \\
&\qquad= \sqrt{\frac{1}{8}(j_1+j_4)(j_2+j_5)(j_3+j_6)}\, \D{j_1}{j_1}{j_1}{h_{e_1}}_x \cdots \D{j_6}{j_6}{j_6}{h_{e_6}}_z + {\cal O}\bigl(\sqrt j\bigr). \label{V_v*6node}
\end{align}
Here the leading term agrees with the diagonal action of the reduced volume operator, as given \eg in \cite{1506b}.

\subsection{Hamiltonian}

\begin{figure}[t]
	\centering
		\includegraphics[width=0.4\textwidth]{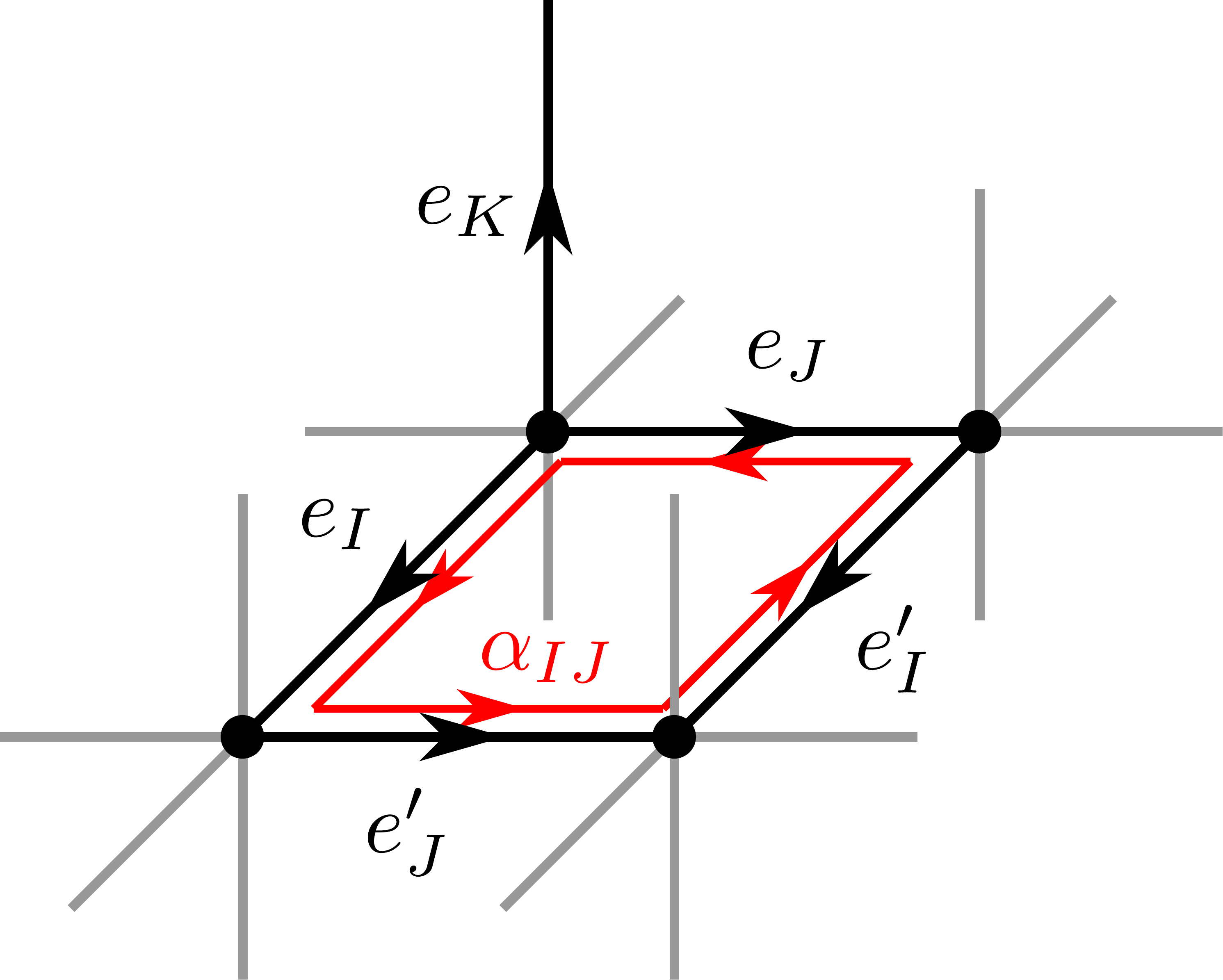}
		\caption{A graph-preserving regularization of Thiemann's Hamiltonian.}
		\label{fig:loop}
\end{figure}

In the literature of the quantum-reduced model, the operator governing the dynamics of the model has usually been taken as a particular version of the Euclidean part of Thiemann's Hamiltonian constraint operator \cite{QSD}. When acting on the node $v$ of a cylindrical function, Thiemann's Hamiltonian is essentially the operator
\be\label{H_E}
H_E^{(v)} = \sum_{\substack{(e_I,e_J,e_K) \\ \text{at $v$}}} \epsilon^{IJK}\,{\rm Tr}\,\Bigl(D^{(s)}(h_{\alpha_{IJ}})D^{(s)}(h_{s_K}^{-1})V_vD^{(s)}(h_{s_K})\Bigr).
\ee
Here $s_K$ denotes a segment of the edge $e_K$, and $\alpha_{IJ}$ is a closed loop associated to the pair of edges ($e_I$, $e_J$). In Thiemann's original formulation, $\alpha_{IJ}$ is a triangular loop spanned by the segments $s_I$ and $s_J$, but in the quantum-reduced model one considers a graph-preserving regularization of the Hamiltonian, which was introduced by Thiemann in \cite{phoenix} to construct the so-called master constraint operator. When the graph-preserving regularization is adapted to a cuboidal graph, the ''segment'' $s_K$ coincides with the edge $e_K$, and the loop $\alpha_{IJ}$ is a rectangular loop formed by the edges $e_I$ and $e_J$, and by two neighboring edges of the reduced spin network state, as illustrated in \Fig{fig:loop}.

The results found for the holonomy operator in section \ref{sec:holonomy} imply that when the Hamiltonian acts on a reduced spin network state, the contribution of highest order in $j$ arises from the terms containing diagonal matrix elements of the holonomies involved in \Eq{H_E}, when each holonomy is expressed in the basis associated with the direction of the corresponding edge. In order to show how the leading terms in the action of the Hamiltonian can be extracted, let us focus on a single term of the sum in \Eq{H_E},
\be
{\rm Tr}\,\Bigl(D^{(s)}(h_{\alpha_{12}})D^{(s)}(h_{e_3}^{-1})V_vD^{(s)}(h_{e_3})\Bigr),
\ee
assuming that the edges $e_1$, $e_2$ and $e_3$ are oriented respectively along the directions $x$, $y$ and $z$. Expanding the trace in the standard basis in which $J_z$ is diagonal, and discarding the terms which contain off-diagonal elements of the matrices $D^{(j)}(h_{e_3})$ and $D^{(j)}(h_{e_3}^{-1})$, we get\footnote{Repeated indices are not summed over in this section, unless indicated by an explicit summation sign.}
\begin{align}
{\rm Tr}\,\Bigl(D^{(s)}(h_{\alpha_{12}})D^{(s)}(h_{e_3}^{-1})V_vD^{(s)}(h_{e_3})\Bigr) = \sum_m{}&{}\D{s}{m}{m}{h_{\alpha_{12}}}\D{s}{m}{m}{h_{e_3}^{-1}}V_v\D{s}{m}{m}{h_{e_3}} \notag \\
&{}+{}\text{off-diagonal terms}\label{H-expanded}
\end{align}

\noindent Since the loop $\alpha_{12}$ is composed of edges aligned in the $x$- and $y$-directions, the matrix elements $\D{s}{m}{m}{h_{\alpha_{12}}}$ are diagonal with respect to the wrong basis. Breaking down the holonomy around the loop as 
\be
h_{\alpha_{12}} = h_{e_2}^{-1}h_{e_1'}^{-1}h_{e_2'}h_{e_1},
\ee
we may write
\be\label{Ds_mm}
\D{s}{m}{m}{h_{\alpha_{12}}} = \sum_{m'nn'} \Db{s}{m}{m'}{h_{e_2}^{-1}}\Db{s}{m'}{n}{h_{e_1'}^{-1}}\D{s}{n}{n'}{h_{e_2'}}\D{s}{n'}{m}{h_{e_1}}.
\ee
We must now express each holonomy in the basis appropriate to the direction of the corresponding edge, and then pick out the terms involving diagonal matrix elements with respect to the new basis. In order to transform the holonomy $\D{s}{m}{n}{h_{e_i}}$ to the $i$-basis, recall that the states $\ket{sm}_i$ are defined as $\ket{sm}_i = D^{(s)}(g_i)\ket{sm}$, where $g_i$ is a rotation which rotates the $z$-axis into the $i$-axis (see section \ref{sec:xystates} of the Appendix). It follows that
\be
\ket{sm} = \sum_n \D{s}{n}{m}{g_i^{-1}} \ket{sn}_i
\ee
and
\be
\D{s}{m}{n}{h_{e_i}} = \sum_{m'} \D{s}{m}{m'}{g_i}\D{s}{m'}{n}{g_i^{-1}}\D{s}{m'}{m'}{h_{e_i}}_i + \text{off-diagonal terms}
\ee
Using this in \Eq{Ds_mm}, and neglecting the off-diagonal terms, we find
\begin{align}
\D{s}{m}{m}{h_{\alpha_{12}}} = \sum_{m'nn'\mu} &\D{s}{m}{m'}{g_y}\Db{s}{m'}{n}{g_y^{-1}g_x}\Db{s}{n}{n'}{g_x^{-1}g_y}\Db{s}{n'}{\mu}{g_y^{-1}g_x}\D{s}{\mu}{m}{g_x^{-1}} \notag \\
&\times\Db{s}{m'}{m'}{h_{e_2}^{-1}}_y\Db{s}{n}{n}{h_{e_1'}^{-1}}_x\D{s}{n'}{n'}{h_{e_2'}}_y\D{s}{\mu}{\mu}{h_{e_1}}_x \label{Dmm leading}
\end{align}
We now obtain the result of our calculation by combining \Eq{Dmm leading} with \Eq{H-expanded}. In order to express the result in a more compact form, we introduce the formal matrix notation
\be\label{bfD}
{\bf D}^{(s)}(h_e)_i = \begin{pmatrix} \D{s}{s}{s}{h_e}_i & & & \\ & \D{s}{s-1}{\;s-1}{h_e}_i & & \\ & & \ddots & \\ & & & & \D{s}{-s}{\;-s}{h_e}_i \end{pmatrix},
\ee
where the matrix has non-zero entries only on the diagonal, and in terms of which we may state our conclusion as follows: The leading term in the action of the operator
\be
{\rm Tr}\,\Bigl(D^{(s)}(h_{\alpha_{12}})D^{(s)}(h_{s_3}^{-1})V_vD^{(s)}(h_{s_3})\Bigr)
\ee
on a reduced spin network state can be found by replacing the operator with
\begin{align}
{\rm Tr}\,\Bigl(&D^{(s)}(g_y) {\bf D}^{(s)}\bigl({h_{e_2}^{-1}}\bigr)_y D^{(s)}\bigl(g_y^{-1}g_x\bigr) {\bf D}^{(s)}\bigl(h_{e_1'}^{-1}\bigr)_x D^{(s)}\bigl(g_x^{-1}g_y\bigr) \notag \\
&\times {\bf D}^{(s)}(h_{e_2'})_y D^{(s)}\bigl(g_y^{-1}g_x\bigr) {\bf D}^{(s)}(h_{e_1})_x D^{(s)}(g_x^{-1}) {\bf D}^{(s)}\bigl({h_{e_3}^{-1}}\bigr)_z V_v {\bf D}^{(s)}(h_{e_3})_z\Bigr). \label{H-final}
\end{align}
The operator \eqref{H-final} involves only the diagonal matrix elements of each holonomy with respect to the appropriate basis. 

In all earlier work concerning the Hamiltonian in the quantum-reduced model, the holonomies involved in the Hamiltonian have been taken in the fundamental representation. When $s=1/2$, the operator \eqref{H-final} agrees with the reduced Hamiltonian discussed \eg in \cite{1402} and \cite{1506a}. In this case, our treatment of the Hamiltonian contains no fundamentally new features at the technical level; only the interpretation of the calculation is different. The Hamiltonian in the quantum-reduced model has previously been viewed as a ''reduced'' operator, obtained by taking the expression \eqref{H_E} for the Hamiltonian in the full theory, and replacing the holonomy operators and the volume operator with their reduced counterparts. In contrast, here we considered the Hamiltonian as an operator in the full theory, and looked for the terms of highest order in $j$ in the action of the operator on a reduced spin network state. When $s=1/2$, both approaches lead to the same result. However, for higher values of $s$ the two points of view are not equivalent, since all the diagonal matrix elements of the holonomy operator (and not only those having a maximal or minimal value of the magnetic index) are involved in the operator \eqref{H-final}, and all of them contribute to the action of the operator at leading order in $j$.

\section{Conclusions}

In this work we considered the operators of quantum-reduced loop gravity from the perspective of full loop quantum gravity. We demonstrated that when the operators of the full theory act on states in the Hilbert space of the quantum-reduced model, the term of leading order in $j$ reproduces the action of the corresponding quantum-reduced operator. Since the ''reduced spin network states'' of the quantum-reduced model are assumed to carry large spins on each of their edges, discarding the terms of lower order in $j$ is well justified.

In the literature of quantum-reduced loop gravity, the operators of the quantum-reduced model are introduced as projections of the operators of the full theory down to the reduced Hilbert space. However, our calculations show that despite their considerable simplicity, the ''reduced'' operators are simply the operators of the full theory acting on states in the Hilbert space of the quantum-reduced model, which is a subspace of the kinematical Hilbert space of the full theory\footnote{Alternatively, if one still prefers to think of the reduced operators as projections of the full theory operators, we have shown that the terms which get projected out are negligibly small in comparison with the terms which are preserved by the projection.}. This result clarifies an important aspect of the relation between the quantum-reduced model and proper loop quantum gravity. 

The relation between the quantum-reduced model and the full theory has been previously discussed, in particular, in the article \cite{1309}. Specifically, it is shown in \cite{1309} that the quantum-reduced model arises from a quantum gauge-fixing to a diagonal triad in the kinematical Hilbert space of full loop quantum gravity. Hence the article establishes the important fact that the quantum-reduced framework is considerably more general than initially thought: It is relevant not only in the context of cosmology, but is applicable to any spatial metric which can be made diagonal by a gauge fixing. On the other hand, in \cite{1309} the quantum-reduced operators are still understood as operators of the full theory projected onto the Hilbert space of the quantum-reduced model. By showing that the projection is irrelevant at leading order in $j$, so that the operators of the quantum-reduced model can be seen simply as the operators of the full theory without the need to introduce any projection, we therefore uncover a different and previously unknown facet of the relation between the quantum-reduced model and full loop quantum gravity.

Our findings also strengthen the technical foundations on which the kinematical framework of quantum-reduced loop gravity is based, since they show that the only essential assumption required to obtain the kinematical structure of the quantum-reduced model is the form of the reduced Hilbert space. If one accepts the reduced Hilbert space as given, then the rest of the quantum-reduced kinematics -- namely, the very simple ''reduced'' operators -- are obtained simply by taking the operators of full loop quantum gravity and letting them act on states in the reduced Hilbert space. One only has to keep in mind that the spin quantum numbers carried by the states in the reduced Hilbert space are assumed to be large, and neglect terms which are of lower than leading order in $j$.

In the case of the holonomy operator, our calculations revealed a modified version of the reduced recoupling rule, which defines the action of the reduced holonomy operator on states in the reduced Hilbert space. We found that all the diagonal components of the operator $\D{j}{m}{n}{h_e}$ act as valid quantum-reduced operators, whereas in the standard formulation of quantum-reduced loop gravity, only the components labeled with the maximal or the minimal value of the magnetic index, \ie $m=n=\pm j$, are taken into account. For $j=1/2$ there is no difference between the two versions of the reduced holonomy operator, and holonomy operators carrying a spin higher than $1/2$ have, to the author's best knowledge, not been used in concrete calculations in the literature of the quantum-reduced model so far. In order to investigate which version of the reduced holonomy operator is physically correct, one could repeat some calculation which has already been carried out in the quantum-reduced model -- for instance, the semiclassical analysis of the dynamics performed in \cite{1402} -- using a Hamiltonian which has been regularized in terms of holonomies carrying a spin higher than $1/2$. One would expect to find that not both versions of the reduced holonomy operator lead to the correct semiclassical limit of the dynamics.

As a more speculative outlook, our results seem to suggest that, in some sense, the quantum-reduced model could be seen as the leading term in a large-$j$ expansion of (a particular sector of) the full theory. It could be worth while to look for a way to turn this intuitive idea into a precise statement, by giving a proper definition of the hypothetical large-$j$ expansion. Taking the expansion to higher orders in $1/j$ would then presumably provide a systematic scheme for refining the approximation encapsulated in the quantum-reduced model. Under such an approach, one would possibly have to re-examine the physical interpretation of the quantum-reduced model, since it is not clear whether the entire scheme, including the subleading terms of the expansion, could still be interpreted as the quantum realization of a particular classical gauge fixing.

\vspace{12pt}

\begin{center}
\large{\bf{Acknowledgments}}
\end{center}

\noindent The author thanks Mehdi Assanioussi for comments on the manuscript. This work was supported by the Polish National Science Centre grant no. 2018/30/Q/ST2/00811.

\appendix

\section{$SU(2)$ and angular momentum}\label{app:SU2}

In this section we collect a number of elementary results from the representation theory of $SU(2)$ and the quantum theory of angular momentum, which are used in the calculations carried out in the main part of this article. The purpose of this section is, above all, to provide a full disclosure of our notation and conventions. A more complete discussion of the material presented below can be found in any textbook of the quantum theory of angular momentum, for example \cite{BrinkSatchler}. In addition, the book by Varshalovich et al. \cite{Varshalovich} provides an encyclopedic collection of formulas related to quantum angular momentum, including, in particular, all the explicit expressions for the Clebsch--Gordan coefficients invoked throughout the calculations performed in section \ref{sec:holonomy}.

\subsection{Fundamental representation}

The group $SU(2)$ consists of $2\times 2$ -matrices of the form
\be\label{gAB}
g\updown{A}{B} = \begin{pmatrix} \alpha&\beta \\ -\bar\beta &\bar\alpha \end{pmatrix}\qquad \text{where} \qquad |\alpha|^2+|\beta|^2=1.
\ee
The fundamental representation of the group is realized by the action of the matrices $g\updown{A}{B}$ on the space ${\cal H}_{1/2} \cong \C^2$, spanned by the two vectors
\be
\ket + = \begin{pmatrix} 1\\0 \end{pmatrix}, \qquad \ket - = \begin{pmatrix} 0\\1 \end{pmatrix}.
\ee
The antisymmetric tensors
\be
\epsilon_{AB} = \begin{pmatrix} 0&1 \\ -1&0 \end{pmatrix}, \qquad \epsilon^{AB} = \begin{pmatrix} 0&1 \\ -1&0 \end{pmatrix}
\ee
are invariant under the action of $SU(2)$:
\be
\epsilon_{AB}g\updown{A}{C}g\updown{B}{D} = \epsilon_{CD}
\ee
and similarly for $\epsilon^{AB}$. By manipulating this relation, one finds that the matrix elements of the inverse matrix $g^{-1}$ are given by
\be
(g^{-1})\updown{A}{B} = \epsilon^{AC}\epsilon_{BD}g\updown{D}{C}.
\ee
Introducing the Pauli matrices
\be
\sigma_x = \begin{pmatrix} 0&1 \\ 1&0 \end{pmatrix}, \qquad \sigma_y = \begin{pmatrix} 0&-i \\ i&0 \end{pmatrix}, \qquad \sigma_z = \begin{pmatrix} 1&0 \\ 0&-1 \end{pmatrix},
\ee
a general element of $SU(2)$ can be parametrized in terms of an angle $\theta$ and a unit vector $\vec n$ as
\be\label{g_n(a)}
g(\theta,\vec n) = e^{-i\theta\vec n\cdot\vec\sigma/2} = \cos\frac{\theta}{2} - i\sin\frac{\theta}{2}(\vec n\cdot\vec\sigma),
\ee
which suggests an interpretation of the group element $g(\theta,\vec n)$ as representing a rotation by the angle $\theta$ around the axis $\vec n$.

\subsection{The angular momentum operator}

The commutator between two Pauli matrices is given by $[\sigma_i,\sigma_j] = 2i\epsilon\downup{ij}{k}\sigma_k$, which implies that the components of the operator $\vec J = \vec\sigma/2$ satisfy
\be\label{J,J}
[J_i,J_j] = i\epsilon\downup{ij}{k}J_k.
\ee
In quantum mechanics, any Hermitian vector operator whose components satisfy the commutation relation \eqref{J,J} is called an angular momentum operator. The commutator \eqref{J,J} encodes the geometric significance of the angular momentum operator as a generator of rotations in three-dimensional space.

All components of $\vec J$ commute with the squared angular momentum
\be
J^2 = J_x^2 + J_y^2 + J_z^2.
\ee
Therefore one can simultaneously diagonalize $J^2$ and one of the components, conventionally chosen as $J_z$. A standard calculation, which follows entirely from the commutation relation \eqref{J,J}, shows that the eigenstates of $J^2$ and $J_z$ obey the eigenvalue equations
\begin{align}
J^2\ket{jm} &= j(j+1)\ket{jm}, \label{J2jm} \\
J_z\ket{jm} &= m\ket{jm}, \label{Jzjm}
\end{align}
where $j$ may be any (positive) integer or half-integer, and $m$ ranges from $-j$ to $j$ in steps of $1$. In the process of the calculation one finds that the operators
\be\label{Jpm-def}
J_\pm = J_x\pm iJ_y
\ee
raise and lower the eigenvalue of $J_z$ by one, while leaving the eigenvalue of $J^2$ unchanged:
\be\label{Jpm}
J_\pm\ket{jm} = \sqrt{j(j+1)-m(m\pm 1)}\ket{j,m\pm 1}.
\ee
In principle, the right-hand side of \Eq{Jpm} contains an arbitrary phase factor, which is not determined by the commutation relation \eqref{J,J}. In this article we follow the Condon--Shortley phase convention, according to which this factor is set equal to $+1$.

\subsection{Spin-$j$ representation}

For a given value of $j$, the states $\ket{jm}$ span the $(2j+1)$-dimensional vector space ${\cal H}_j$, as the index $m$ takes the values $-j,-j+1,\dots,j$. The spin-$j$ representation of $SU(2)$ is defined by the matrices representing the operators $g(\theta,\vec n) = e^{-i\theta\vec n\cdot\vec J}$ on the space ${\cal H}_j$. These matrices, whose matrix elements are given by
\be
\D{j}{m}{n}{g} = \bra{jm}e^{-i\theta\vec n\cdot\vec J}\ket{jn},
\ee
are known as the Wigner matrices. We adopt the definition
\be
\epsilon^{(j)}_{mn} = (-1)^{j-m}\delta_{m,-n}
\ee
for the invariant epsilon tensor in the spin-$j$ representation. Then the inverse matrix $D^{(j)}(g^{-1})$ is given by the relation
\be\label{Dj-inv}
\D{j}{m}{n}{g^{-1}} = \epsilon^{(j)}_{mm'}\epsilon^{(j)}_{nn'}\D{j}{n'}{m'}{g} = (-1)^{m-n}\D{j}{-n\;}{-m}{g}
\ee
By analogy with the definition $\tau_i = -i\sigma_i/2$ in the fundamental representation, we define the anti-Hermitian generators of $SU(2)$ in the spin-$j$ representation as
\be\label{tau^j}
\Tau{j}{i}{m}{n} = -i\bra{jm}J_i\ket{jn}.
\ee
Note that the matrix elements of the generators are given in explicit form by \Eqs{Jzjm}--\eqref{Jpm}.

In order to clarify how the spin-$j$ representation of $SU(2)$ is related to the fundamental representation, let us consider the $2j$-fold tensor product state
\be\label{Psi_j}
\ket{\Psi_j} = \ket +\otimes\ket + \otimes \cdots \otimes\ket +.
\ee
By direct calculation, one finds that the state \eqref{Psi_j} is an eigenstate of the total angular momentum operator
\be
J^{({\rm tot})} = J^{(1)} + J^{(2)} + \dots + J^{(2j)}
\ee
(where each $J^{(i)}$ acts on the $i$-th factor of the tensor product $\otimes_{i=1}^{2j} {\cal H}_{1/2}^{(i)}$), as indicated by the eigenvalue equations\footnote{The eigenvalue equation for the $z$-component of $J^{({\rm tot})}$ is immediate. In order to verify the equation for $(J^{({\rm tot})})^2$, note that
\[
\bigl(J^{({\rm tot})}\bigr)^2 = \sum_i \bigl(J^{(i)}\bigr)^2 + \sum_{i\neq k} \vec J^{\,(i)}\cdot\vec J^{\,(k)},
\]
where the cross terms can be written as
\[
\vec J^{\,(i)}\cdot\vec J^{\,(k)} = J^{(i)}_z J^{(k)}_z + \frac{1}{2}\bigl(J^{(i)}_+ J^{(k)}_- + J^{(i)}_- J^{(k)}_+\bigr).
\]
When acting on the state \eqref{Psi_j}, only the terms $(J^{(i)})^2$ and $J^{(i)}_z J^{(k)}_z$ give a non-vanishing result, since the state $\ket +$ is annihilated by the raising operator $J_+$.
}
\begin{align}
\bigl(J^{({\rm tot})}\bigr)^2\ket{\Psi_j} &= j(j+1)\ket{\Psi_j}, \\
J^{({\rm tot})}_z\ket{\Psi_j} &= j\ket{\Psi_j}.
\end{align}
This shows that the state \eqref{Psi_j} can be identified as the state $\ket{jj}$. The remaining states $\ket{jm}$ are then obtained by repeatedly applying the lowering operator
\be
J_-^{({\rm tot})} = J_-^{(1)} + J_-^{(2)} + \dots + J_-^{(2j)}.
\ee
In this way one finds
\be\label{jm}
\ket{jm} = \sqrt{\frac{(j+m)!(j-m)!}{(2j)!}}\Bigl(\underbrace{\ket +\otimes\cdots\otimes\ket +}_{\text{$j+m$ times}} \otimes \underbrace{\ket -\otimes \cdots \otimes\ket -}_{\text{$j-m$ times}}\;+\;\text{all permutations}\Bigr),
\ee
establishing a direct relation between the spaces ${\cal H}_j$ and ${\cal H}_{1/2}$. Indeed, the spin-$j$ representation of $SU(2)$ is often introduced in the literature in terms of the completely symmetric subspace of the $2j$-fold tensor product space ${\cal H}_{1/2}\otimes\cdots\otimes{\cal H}_{1/2}$.

\subsection{Clebsch--Gordan coefficients}

The tensor product space ${\cal H}_{j_1}\otimes{\cal H}_{j_2}$ is spanned by the states $\ket{j_1m_1}\ket{j_2m_2}$, which are eigenstates of the mutually commuting operators
\be
\bigl(J^{(1)}\bigr)^2, \qquad \bigl(J^{(2)}\bigr)^2, \qquad J^{(1)}_z, \qquad J^{(2)}_z.
\ee
The operators
\be
\bigl(J^{(1)}\bigr)^2, \qquad \bigl(J^{(2)}\bigr)^2, \qquad \bigl(J^{(1)}+J^{(2)}\bigr)^2, \qquad J^{(1)}_z+J^{(2)}_z
\ee
form another complete set of commuting operators on ${\cal H}_{j_1}\otimes{\cal H}_{j_2}$. Let us denote their eigenstates by $\ket{j_1j_2;jm}$. Since both sets of states provide a basis of ${\cal H}_{j_1}\otimes{\cal H}_{j_2}$, they must be related to each other by unitary transformations of the form
\be\label{uncoupled}
\ket{j_1m_1}\ket{j_2m_2} = \sum_{jm} \CG{j_1}{j_2}{j}{m_1}{m_2}{m}\ket{j_1j_2;jm}
\ee
and
\be\label{coupled}
\ket{j_1j_2;jm} = \sum_{m_1m_2} \CGi{j_1}{j_2}{j}{m_1}{m_2}{m}\ket{j_1m_1}\ket{j_2m_2}.
\ee
The coefficients in these expansions are known as the Clebsch--Gordan coefficients.

The Condon--Shortley phase convention fixes the phases of the Clebsch--Gordan coefficients by the requirement that the coefficient $\CG{j_1}{j_2}{j}{j_1}{j-j_1}{j}$ is real and positive, and by the phase choice made in \Eq{Jpm}. Under the Condon--Shortley convention, all of the Clebsch--Gordan coefficients are real-valued. Then the coefficient $\braket{j_1j_2;jm}{j_1m_1\otimes j_2m_2}$ is numerically equal to the inverse coefficient $\braket{j_1m_1\otimes j_2m_2}{j_1j_2;jm}$. For this reason, the coefficients appearing in \Eq{uncoupled} are usually not distinguished from the inverse coefficients appearing in \Eq{coupled}.

Some basic properties of the Clebsch--Gordan coefficients follow immediately from their definition. The coefficient $\CG{j_1}{j_2}{j}{m_1}{m_2}{m}$ vanishes unless the conditions
\be\label{CGtriangle}
|j_1-j_2| \leq j \leq j_1+j_2 \qquad \text{and} \qquad j_1+j_2+j = {\rm integer}
\ee
as well as
\be\label{m1+m2}
m= m_1+m_2
\ee
are met. Moreover, the Clebsch--Gordan coefficients satisfy the orthogonality relations
\be\label{CG-orth-jm}
\sum_{jm} \CG{j_1}{j_2}{j}{m_1}{m_2}{m}\CGi{j_1}{j_2}{j}{m_1'}{m_2'}{m} = \delta_{m_1m_1'}\delta_{m_2m_2'}
\ee
and
\be\label{CG-orth-mm}
\sum_{m_1m_2} \CG{j_1}{j_2}{j}{m_1}{m_2}{m}\CGi{j_1}{j_2}{j'}{m_1}{m_2}{m'} = \delta_{jj'}\delta_{mm'}.
\ee
By applying an $SU(2)$ rotation to \Eq{uncoupled}, one can deduce the Clebsch--Gordan series
\be\label{CG-ser}
\D{j_1}{m_1}{n_1}{g}\D{j_2}{m_2}{n_2}{g} = \sum_{jmn} \CGi{j_1}{j_2}{j}{m_1}{m_2}{m}\CG{j_1}{j_2}{j}{n_1}{n_2}{n}\D{j}{m}{n}{g}
\ee
for the matrix elements of the Wigner matrices. Recalling the condition \eqref{m1+m2}, we may eliminate the sums over $m$ and $n$, and write the Clebsch--Gordan series in the equivalent form
\be\label{CG-ser-j}
\D{j_1}{m_1}{n_1}{g}\D{j_2}{m_2}{n_2}{g} = \sum_{j} \CGi{j_1}{j_2}{j}{m_1}{m_2}{m_1+m_2}\CG{j_1}{j_2}{j}{n_1}{n_2}{n_1+n_2}\D{j}{m_1+m_2\;}{n_1+n_2}{g}.
\ee

\subsection{Eigenstates of $J_x$ and $J_y$}\label{sec:xystates}

The states $\ket{jm}_i$, which diagonalize the operators $J^2$ and $J_i$ (for $i=x$ or $y$), can be constructed by starting with the states $\ket{jm}$ and applying a rotation which rotates the $z$-axis into the $x$-axis or the $y$-axis. If $g_i$ is an $SU(2)$ element representing any rotation which rotates the vector $\hat e_z$ into the vector $\hat e_i$, the state
\be
\ket{jm}_i = D^{(j)}(g_i)\ket{jm}
\ee
is an eigenstate of the operators $J^2$ and $J_i$ with eigenvalues $j(j+1)$ and $m$.

The group element $g_i$ is not uniquely determined by the requirement that the corresponding rotation must rotate the $z$-axis into the $i$-axis. However, the diagonal matrix elements of the Wigner matrices in the basis $\ket{jm}_i$,
\be\label{Dmm_i}
\D{j}{m}{m}{g}_i = {}_i\bra{jm}D^{(j)}(g)\ket{jm}_i,
\ee
which play a central role in this article, are independent of the choice of rotation used to construct the states $\ket{jm}_i$. To verify this, note that if $g_i$ and $g_i'$ are group elements describing two different rotations which rotate the $z$-axis into the $i$-axis, the combined rotation $g_i'g_i^{-1}$ preserves the $i$-axis, so it must have the form $g_i'g_i^{-1} = e^{i\alpha\sigma_i}$. In other words, $g_i' = e^{i\alpha\sigma_i}g_i$, which implies that the states $\ket{jm}_i'$, constructed by applying the rotation $g_i'$ to the states $\ket{jm}$, are related by phase factors to the states constructed using the rotation $g_i$:
\be
\ket{jm}_i' = e^{i\beta(j,m)}\ket{jm}_i.
\ee
(This is also clear from the fact that both $\ket{jm}_i$ and $\ket{jm}_i'$ are eigenstates of the operators $J^2$ and $J_i$ corresponding to the non-degenerate pair of eigenvalues $j(j+1)$ and $m$.) When the diagonal matrix elements of $D^{(j)}(g)$ are taken in the basis $\ket{jm}_i'$, the phase factors $e^{i\beta(j,m)}$ cancel, so the diagonal matrix elements indeed do not depend on which rotation is selected to construct the basis $\ket{jm}_i$.

Whenever an explicit choice of the rotation $g_i$ has to be made, we will choose a rotation corresponding to a cyclic permutation of the coordinate axes, \ie a rotation which rotates the axes $(x,y,z)$ into $(y,z,x)$ or $(z,x,y)$. This choice has the advantage that the action of the angular momentum operator on the states $\ket{jm}_i$ is particularly easy to deduce. \Eqs{Jzjm}--\eqref{Jpm} show that the components of the angular momentum operator act on the states $\ket{jm} \equiv \ket{jm}_z$ as
\begin{align}
J_x\ket{jm}_z &= C_+(j,m)\ket{j,m+1}_z + C_-(j,m)\ket{j,m-1}_z \label{Jx*z} \\
J_y\ket{jm}_z &= -iC_+(j,m)\ket{j,m+1}_z +iC_-(j,m)\ket{j,m-1}_z \label{Jy*z} \\
J_z\ket{jm}_z &= m\ket{jm}_z \label{Jz*z}
\end{align}
where we have introduced the abbreviation
\be
C_\pm(j,m) = \frac{1}{2}\sqrt{j(j+1)-m(m\pm 1)}.
\ee
If the states $\ket{jm}_i$ are defined in the way described above, we may cyclically permute the labels $x$, $y$ and $z$ to find
\begin{align}
J_x\ket{jm}_x &= m\ket{jm}_x \label{Jx*x} \\
J_y\ket{jm}_x &= C_+(j,m)\ket{j,m+1}_x + C_-(j,m)\ket{j,m-1}_x \label{Jy*x} \\
J_z\ket{jm}_x &= -iC_+(j,m)\ket{j,m+1}_x +iC_-(j,m)\ket{j,m-1}_x \label{Jz*x}
\end{align}
and
\begin{align}
J_x\ket{jm}_y &= -iC_+(j,m)\ket{j,m+1}_y +iC_-(j,m)\ket{j,m-1}_y \label{Jx*y} \\
J_y\ket{jm}_y &= m\ket{jm}_y \label{Jy*y} \\
J_z\ket{jm}_y &= C_+(j,m)\ket{j,m+1}_y + C_-(j,m)\ket{j,m-1}_y \label{Jz*y}
\end{align}

\section{Square root of the operator $q_v$}\label{app:sqrt}

In section \ref{sec:volume} we encountered the problem of computing the action of the volume operator $V_v = \sqrt{|q_v|}$ on the reduced spin network node \eqref{6-node}, given that the action of the operator $q_v$ on the node is approximately diagonal, as shown by \Eq{q_v*6node}. Here we will present a detailed solution of this problem. The solution is based on treating the off-diagonal terms in \Eq{q_v*6node} as a perturbation over the diagonal term, and using standard perturbation theory to extract the leading term (as well as the first subleading terms) in the action of $\sqrt{|q_v|}$ on the state \eqref{6-node}. 

In order to carry out the analysis in detail, let us consider the equivalent but notationally lighter problem of finding the action of the operator $\sqrt{|Q|}$, with $Q$ given by
\be\label{simpleQ}
Q = \epsilon^{ijk}\bigl(J_i^{(1)}+J_i^{(4)}\bigr)\bigl(J_j^{(2)}+J_j^{(5)}\bigr)\bigl(J_k^{(3)}+J_k^{(6)}\bigr),
\ee
on the state
\be\label{j-state}
\ket{j_1j_1}_x\ket{j_2j_2}_y\ket{j_3j_3}_z\ket{j_4j_4}_x\ket{j_5j_5}_y\ket{j_6j_6}_z
\ee
in ${\cal H}_{j_1}\otimes\cdots\otimes{\cal H}_{j_6}$. The action of $Q$ on the generic basis state
\be
\ket{j_1m_1}_x\ket{j_2m_2}_y\ket{j_3m_3}_z\ket{j_4m_4}_x\ket{j_5m_5}_y\ket{j_6m_6}_z
\ee
produces the diagonal term
\begin{align}
&\bigl(J_x^{(1)}+J_x^{(4)}\bigr)\bigl(J_y^{(2)}+J_y^{(5)}\bigr)\bigl(J_z^{(3)}+J_z^{(6)}\bigr)\ket{j_1m_1}_x\ket{j_2m_2}_y\ket{j_3m_3}_z\ket{j_4m_4}_x\ket{j_5m_5}_y\ket{j_6m_6}_z \notag \\
&= (m_1+m_4)(m_2+m_5)(m_3+m_6)\ket{j_1m_1}_x\ket{j_2m_2}_y\ket{j_3m_3}_z\ket{j_4m_4}_x\ket{j_5m_5}_y\ket{j_6m_6}_z \label{Qdiag}
\end{align}
as well as off-diagonal terms of the form
\begin{align}
&J_x^{(1)}J_z^{(2)}J_y^{(3)}\ket{j_1m_1}_x\ket{j_2m_2}_y\ket{j_3m_3}_z\ket{j_4m_4}_x\ket{j_5m_5}_y\ket{j_6m_6}_z \notag \\
&=m_1\ket{j_1m_1}\Bigl(C_+(j_2,m_2)\ket{j_2,m_2+1}_y + C_-(j_2,m_2)\ket{j_2,m_2-1}_y\Bigr) \notag \\
&\times\Bigl(-iC_+(j_3,m_3)\ket{j_3,m_3+1}_z + iC_-(j_3,m_3)\ket{j_3,m_3-1}_y\Bigr)\ket{j_4m_4}_x\ket{j_5m_5}_y\ket{j_6m_6}_z \label{Qoff1}
\end{align}
and
\begin{align}
&J_y^{(1)}J_z^{(2)}J_x^{(3)}\ket{j_1m_1}_x\ket{j_2m_2}_y\ket{j_3m_3}_z\ket{j_4m_4}_x\ket{j_5m_5}_y\ket{j_6m_6}_z \notag \\
&=\Bigl(C_+(j_1,m_1)\ket{j_1,m_1+1}_x + C_-(j_1,m_1)\ket{j_1,m_1-1}_x\Bigr) \notag \\
&\times\Bigl(C_+(j_2,m_2)\ket{j_2,m_2+1}_y + C_-(j_2,m_2)\ket{j_2,m_2-1}_y\Bigr) \notag \\
&\times\Bigl(C_+(j_3,m_3)\ket{j_3,m_3+1}_z + C_-(j_3,m_3)\ket{j_3,m_3-1}_z\Bigr) \ket{j_4m_4}_x\ket{j_5m_5}_y\ket{j_6m_6}_z, \label{Qoff2}
\end{align}
together with the similar terms which arise from the remaining combinations of the angular momentum operators in \Eq{simpleQ}. 

\begin{figure}[t]
	\centering
		\includegraphics[width=0.35\textwidth]{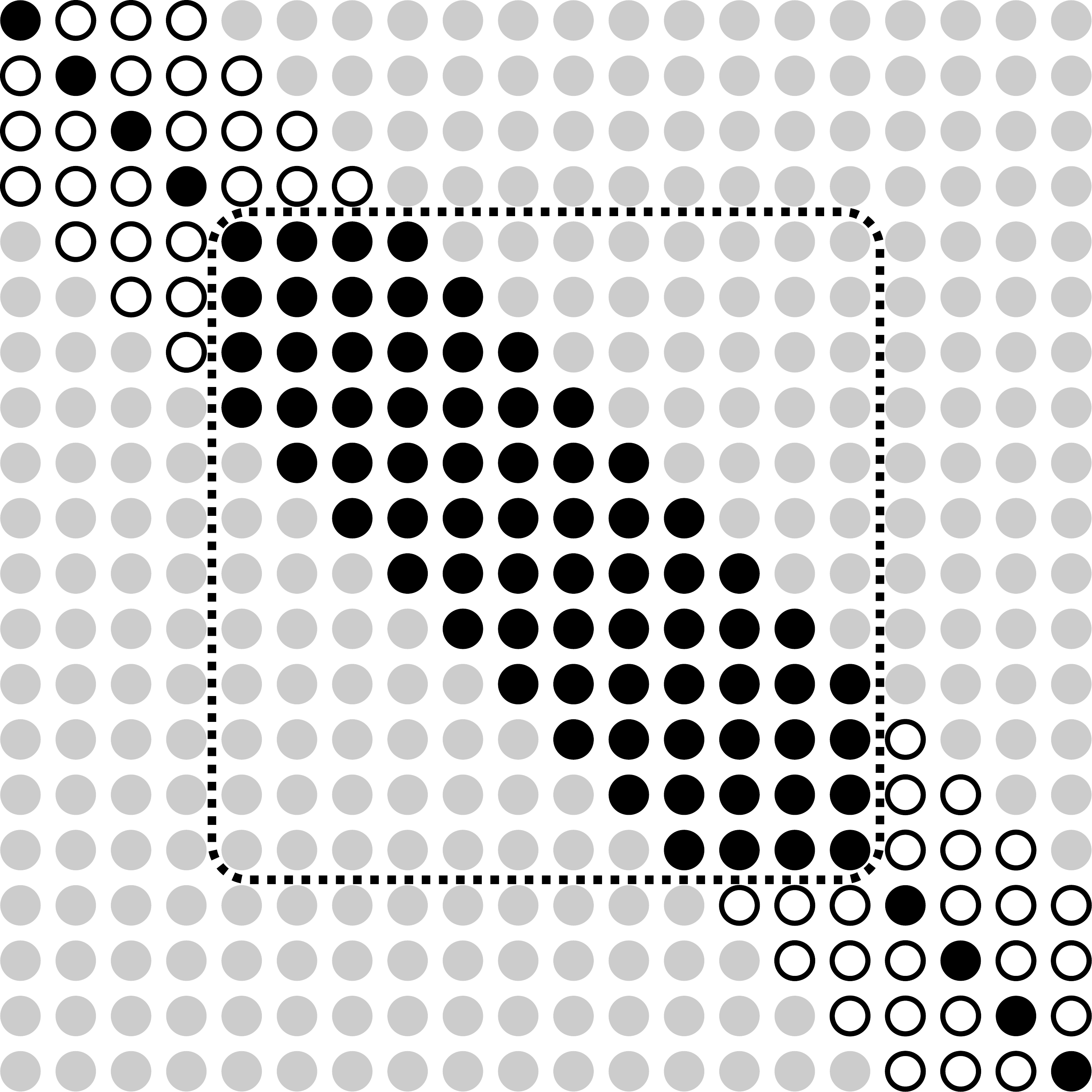}
		\caption{Dividing the matrix $Q$ into an unperturbed part $Q_0$ and a perturbation $W$. Outside of the central block, the diagonal matrix elements are of order $j^3$, while the off-diagonal matrix elements are at most of order $j^2$. Hence the matrix elements marked in white can be considered as a perturbation over the unperturbed matrix formed by the matrix elements marked in black. The matrix elements marked in grey are equal to zero.}
		\label{fig:Q0plusW}
\end{figure}

When $m$ is close enough to $j$ in absolute value, say $j-|m| = {\cal O}(1)$, the coefficient $C_\pm(j,m)$ is of order $\sqrt j$. Thus, within the sector of the space ${\cal H}_{j_1}\otimes\cdots\otimes{\cal H}_{j_6}$ in which each $|m_I|$ is close to the corresponding $j_I$, the off-diagonal matrix elements given by \eqref{Qoff1} and \eqref{Qoff2} are suppressed by at least a factor of $1/j$ in relation to the diagonal matrix elements of \Eq{Qdiag}. Therefore the matrix representing the operator $Q$ can be divided into an unperturbed part $Q_0$ and a small perturbation $W$ in the way indicated by the schematic drawing in \Fig{fig:Q0plusW}. Assume that the rows and the columns of the matrix are labeled so that the diagonal elements are ordered from largest to smallest. We delineate a central block of the matrix in such a way that everywhere outside the block, the diagonal matrix elements are of order $j^3$, while the off-diagonal elements are at most of order $j^2$. The unperturbed matrix $Q_0$ is then defined to consist of the diagonal matrix elements outside the central block, and of all the matrix elements inside the block. The off-diagonal matrix elements outside the central block are assigned to the perturbation $W$. (The exact location of the central block's boundary is irrelevant to our analysis, as long as we are interested in computing the action of $\sqrt{|Q|}$ on the state \eqref{j-state} only.)

Inserting a formal small parameter $\epsilon$ to keep track of powers of the perturbation, we know that perturbation theory can be used to approximate the spectrum of the operator
\be
Q = Q_0 + \epsilon W
\ee
power-by-power in $\epsilon$. The approximation is expressed in terms of the eigenvalues $\lambda_i^{(0)}$ and eigenstates $\ket{\lambda_i^{(0)}}$ of the unperturbed operator $Q_0$, which are assumed to be known. The first-order approximations for the eigenvalues and eigenstates of $Q$ are given by
\be\label{lambda-1}
\lambda_i = \lambda_i^{(0)} + \epsilon W_{ii} + {\cal O}(\epsilon^2)
\ee
and
\be\label{state-1}
\ket{\lambda_i} = \bket{\lambda_i^{(0)}} + \epsilon\;{\sum_{k\neq i}}^\prime \frac{W_{ki}}{\lambda_i^{(0)}-\lambda_k^{(0)}}\,\bket{\lambda_k^{(0)}}  + {\cal O}(\epsilon^2),
\ee
where we have introduced the notation
\be
W_{ik} = \bbra{\lambda_i^{(0)}}W\bket{\lambda_k^{(0)}}
\ee
for the matrix elements of the perturbation in the basis of unperturbed eigenstates. The prime on the sum over $k$ in \Eq{state-1} indicates that whenever some of the unperturbed eigenvalues are degenerate, one should choose the basis of unperturbed eigenstates in such a way that the perturbation $W$ has no non-vanishing matrix elements between different eigenstates corresponding to the same degenerate eigenvalue, and after this has been done, the terms with $\lambda_k^{(0)} = \lambda_i^{(0)}$ are to be excluded from the sum. However, in the present problem there is no need to take this point explicitly into account, since we are only interested in the action of $\sqrt{|Q|}$ on the state \eqref{j-state}, which is a non-degenerate eigenstate of the unperturbed operator $Q_0$.

Applying the approximations \eqref{lambda-1} and \eqref{state-1} to the spectral decomposition of the operator $\sqrt{|Q|}$,
\be\label{sqrtQ-spectr}
\sqrt{|Q|} = \sum_i \sqrt{|\lambda_i|}\,\ket{\lambda_i}\bra{\lambda_i},
\ee
we obtain
\begin{align}
\quad\sqrt{|Q|} = \sum_i \biggl(\sqrt{\bigl|\lambda_i^{(0)}\bigr|} &+ \epsilon\frac{W_{ii}}{2\sqrt{\bigl|\lambda_i^{(0)}\bigr|}}\biggr)\bket{\lambda_i^{(0)}}\bbra{\lambda_i^{(0)}} \notag \\
&+ \epsilon\;{\sum_{ik}}' W_{ik}\frac{\sqrt{\bigl|\lambda_i^{(0)}\bigr|} - \sqrt{\bigl|\lambda_k^{(0)}\bigr|}}{\lambda_i^{(0)}-\lambda_k^{(0)}} \bket{\lambda_i^{(0)}}\bbra{\lambda_k^{(0)}} + {\cal O}(\epsilon^2).\quad\label{sqrtQ-approx}
\end{align}
Let us assign the label $i=0$ to the state \eqref{j-state}. Acting with the operator \eqref{sqrtQ-approx} on this state, we find
\be\label{sqrtQ*lambda0}
\sqrt{|Q|}\,\bket{\lambda_0^{(0)}} = \sqrt{\lambda_0^{(0)}}\,\bket{\lambda_0^{(0)}} + \epsilon\;{\sum_{i\neq 0}} \frac{W_{i0}}{\sqrt{\lambda_i^{(0)}} + \sqrt{\lambda_0^{(0)}}} \bket{\lambda_i^{(0)}} + {\cal O}(\epsilon^2),
\ee
where we have noted that (1) the expectation value $W_{00}$ vanishes; (2) the unperturbed eigenvalue $\lambda_0^{(0)} = (j_1+j_4)(j_2+j_5)(j_3+j_6)$ is positive and non-degenerate; (3) all the eigenvalues $\lambda_i^{(0)}$ corresponding to states for which the matrix element $V_{i0}$ is nonvanishing are also positive; and (4) we have carried out the simplification
\be
\frac{\sqrt{\lambda_i^{(0)}} - \sqrt{\lambda_0^{(0)}}}{\lambda_i^{(0)}-\lambda_0^{(0)}} = \frac{1}{\sqrt{\lambda_i^{(0)}} + \sqrt{\lambda_0^{(0)}}}.
\ee
All the unperturbed eigenvalues $\lambda_i^{(0)}$ entering \Eq{sqrtQ*lambda0} are of order $j^3$, while the matrix elements $W_{i0}$ are at most of order $j^2$. Consequently, the coefficient multiplying the leading term is of order $j^{3/2}$, whereas the coefficients in the sum over $i$ are of order $\sqrt j$. Hence we may remove the formal parameter $\epsilon$ and expect that \eqref{sqrtQ*lambda0} remains a valid approximation for the action of $\sqrt{|Q|}$ on the state \eqref{j-state}. We have therefore shown that
\begin{align}
\sqrt{|Q|}\,\ket{j_1j_1}_x&\cdots\ket{j_6j_6}_z \notag \\
&=\sqrt{(j_1+j_4)(j_2+j_5)(j_3+j_6)}\,\ket{j_1j_1}_x\cdots \ket{j_6j_6}_z + {\cal O}\bigl(\sqrt j\bigr),
\end{align}
which is equivalent to \Eq{V_v*6node} given in the main text for the action of the volume operator on the reduced spin network node \eqref{6-node}.

\renewcommand{\bibname}{References}

\end{document}